\documentclass[9pt,twocolumn,twoside]{osajnl}
\usepackage{color}
\journal{optica}

\setboolean{shortarticle}{false}

\title{Demonstration of a non-Abelian geometric controlled-Not gate in a
superconducting circuit}
\author[1,2]{Kai Xu}
\author[3]{Wen Ning}
\author[3]{Xin-Jie Huang}
\author[3]{Pei-Rong Han}
\author[2]{Hekang Li}
\author[3,4]{Zhen-Biao Yang}
\author[1,2]{Dongning Zheng}
\author[1,2,5]{Heng Fan}
\author[3,6]{Shi-Biao Zheng}

\affil[1]{Institute of Physics and Beijing National Laboratory for Condensed Matter Physics, Chinese Academy of Sciences, Beijing 100190, China}
\affil[2]{CAS Center for Excellence in Topological Quantum Computation, University of Chinese Academy of Sciences, Beijing 100190, China}
\affil[3]{Fujian Key Laboratory of Quantum Information and Quantum Optics,College of Physics and Information Engineering, Fuzhou University, Fuzhou,Fujian 350108, China}
\affil[4]{zbyang@fzu.edu.cn}
\affil[5]{hfan@iphy.ac.cn}
\affil[6]{t96034@fzu.edu.cn}

\begin{abstract}
Holonomies, arising from non-Abelian geometric transformations of quantum
states in Hilbert space, offer a promising way for quantum computation.
These holonomies are not commutable and thus can be used for the realization
of a universal set of quantum logic gates, where the global geometric
feature may result in some noise-resilient advantages. Here we report the
first on-chip realization of a non-Abelian geometric controlled-Not gate in
a superconducting circuit, which is a building block for constructing a
holonomic quantum computer. The conditional dynamics is achieved in an
all-to-all connected architecture involving multiple frequency-tunable
superconducting qubits controllably coupled to a resonator; a holonomic gate
between any two qubits can be implemented by tuning their frequencies on
resonance with the resonator and applying a two-tone drive to one of them.
This gate represents an important step towards the all-geometric realization
of scalable quantum computation on a superconducting platform.
\end{abstract}
 \setboolean{displaycopyright}{true}

\begin{document}

\maketitle

\section{Introduction}
When a nondegenerate quantum system makes a cyclic evolution in the Hilbert
space, it will pick up a phase, which, in general, is contributed by both
the dynamical and geometric effects. The dynamical part is the time integral
of the energy, while the geometric one depends upon the area enclosed by the
loop that the quantum state traverses in the Hilbert space. This effect,
discovered by Berry in cyclic and adiabatic evolutions \cite{Berry_PRSL1984}, has been
generalized to nonadiabatic \cite{Aharonov_PRL1987} and nondegenerate \cite{Wilczek_PRL1984} cases. If a system has
degenerate energy levels, the cyclic evolution of the corresponding
degenerate subspaces will produce a matrix-valued quantum state
transformation that is path-dependent and referred to as non-Abelian
geometric phase or holonomy \cite{Wilczek_PRL1984}. The Berry phase and holonony depend upon
the global geometry of the associated loops and have intrinsic resistance to
certain kinds of small errors, suggesting quantum gates based on geometric
operations have practical advantages as compared to dynamical gates \cite{Chiara_PRL2003,Filipp_PRL2009,Carollo_PRL2003,Zheng_PRA2015}.
In particular, it was shown that all of the elementary one- and two-qubit
gates needed for accomplishing any quantum computation task could be
achieved with Berry phase and holonomic transformations, offering a
possibility for implementations of geometric quantum computation \cite{Zanardi_PLA1999,Pachos_PRA2000}.

The conditional Berry phase was first observed in nuclear magnetic resonance
systems \cite{Jones_Nature2000}. However, the relatively long operation time associated with an
adiabatic evolution represents an unfavorable condition for the
implementation of geometric quantum computation with such controlled phase
gates. As such, geometric effects without the adiabatic restriction are
highly desirable for the implementation of quantum logic gates that are
robust against noises \cite{Wang_PRL2001,Zhu_PRL2002,Nazir_PRA2002,Blais_PRA2003,ZhangJ_PRA2018,Chen_arXiv2012}. So far, nonadiabatic geometric controlled-phase gates have been realized in ion traps \cite{Leibfried_Nature2003,Benhelm_NaturePhysics2008,Ballance_PRL2016,Gaebler_PRL2016} and
superconducting circuits \cite{Song_NatCommun2017,Xu_PRL2020,Xu_PRL2020_Gate}. On the other hand, Sj\"{o}qvist {\it et al}%
. have proposed an approach for realizing a universal set of elementary
gates based on nonadiabatic holonomies \cite{Sjoqvist_NJP2012}, whose robustness against
noises has been analyzed \cite{Johansson_PRA2012,Zheng_PRA2016}. Following this approach, a universal gate
set involving two non-commutable single-qubit gates and a two-qubit
controlled-Not (CNOT) gate have been experimentally realized with nuclear
magnetic resonance \cite{Feng_PRL2013} and solid-state spins \cite{Zu_Nature2014,Arroyo-Camejo_NatCommun2014}. Several groups have
demonstrated holonomic single-qubit gates in superconducting circuits
\cite{Abdumalikov_Nature2013,Xu_PRL2018,Yan_PRL2019,Zhang_arXiv1811}, which represent a promising platform for quantum computation \cite{You_PhysicsToday2005}.
Recently, Egger et al. reported a holonomic operation for producing
entangled states in a superconducting circuit \cite{Egger_PRApplied2019}. 
However, a non-Abelian geometric entangling gate
necessary for constructing a universal holonomic gate set has not been
implemented in such scalable systems. More recently, Han et al. reported a
universal set of time-optimal geometric gates with superconducting qubits
\cite{Han_arXiv2004}, where single-qubit gates were realized using non-Abelian geometric
phase, but the two-qubit gate was based on Abelian geometric phase.

In this paper, we propose and experimentally demonstrate a scheme for
realizing non-adiabatic, non-Abelian geometric CNOT gate for two qubits, one
acting as the control qubit and the other as the target qubit. The two
qubits are strongly coupled to a resonator, so that the energy levels of the
target qubit depend on the state of the control qubit. This conditional
energy-level shift enables the target qubit to be resonantly driven by
classical fields, conditional on the state of the control qubit. With
suitable setting of the parameters, these classical fields can drive the
degenerate subspace spanned by the two basis states of the target qubit to
undergo a conditional cyclic evolution, realizing a CNOT gate between these
two qubits. We realize this holonomic gate in a superconducting multi-qubit
processor, where any two qubits can be selectively coupled to a common
resonator but effectively decoupled from other qubits through frequency
tuning. This flexibility enables direct implementation of holonomic gates
between any pair of qubits on the chip, without the restriction of
nearest-neighbor couplings. The measured process fidelity for the CNOT gate
is above 0.9. With further improvements in the device design and fabrication, as confirmed by our numerical simulations, the gate fidelity can be significantly increased. Our scheme is applicable
to other spin-boson systems, such as cavity QED and ion traps \cite{Buluta_RopiP2011}.

\section{Theoretical model}
\begin{figure}[htbp] 
	\centering
	%\fbox{\includegraphics[width=\linewidth]{Energy_Level.pdf}}
	\includegraphics[width=\linewidth]{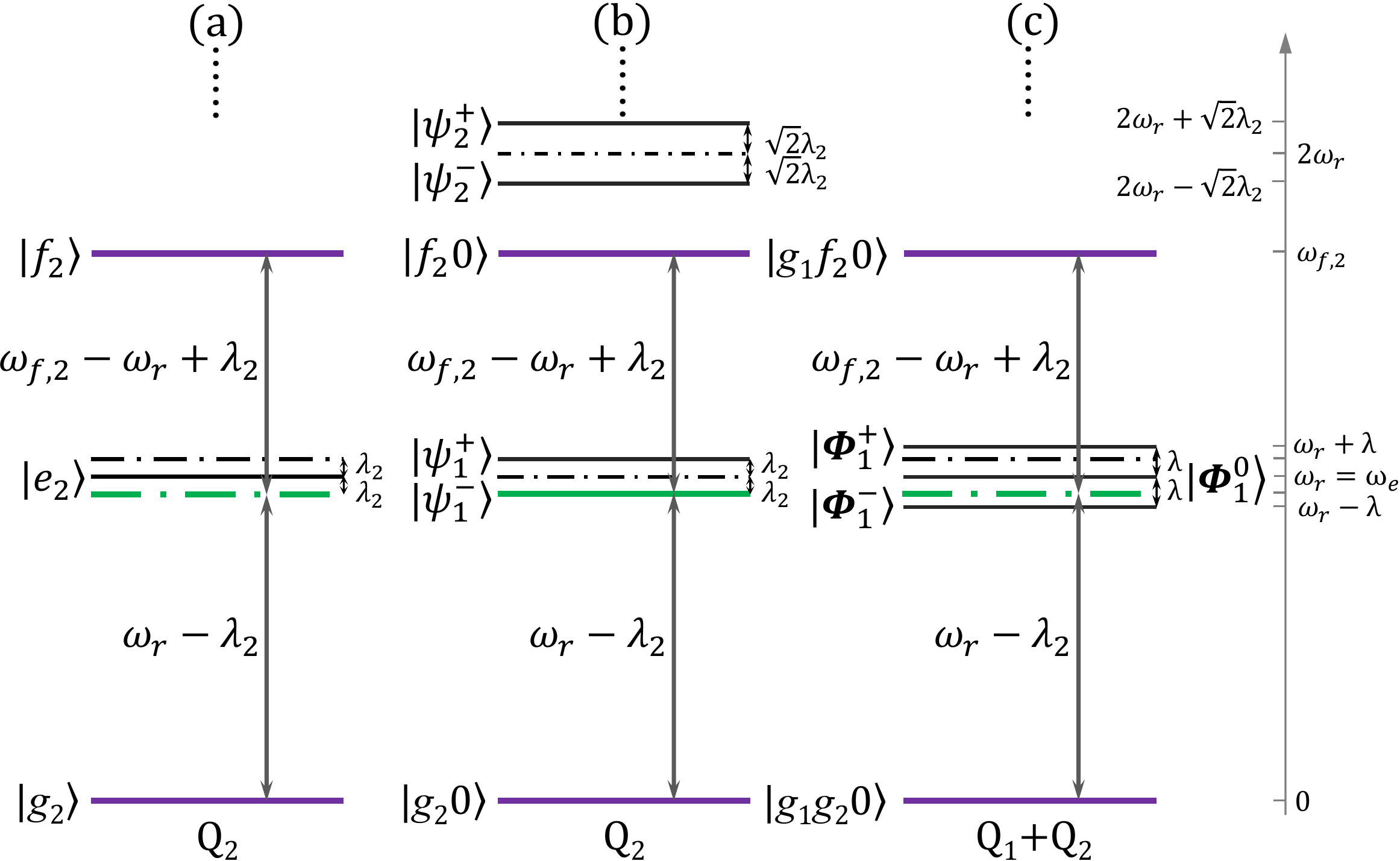}
	\caption{Energy level configuration and excitation scheme for
		the two-qubit CNOT gate. The control and target qubits are denoted as Q$_{1}$ and Q$_{2}$, respectively. (a) Bare energy levels of Q$_{2}$ and frequencies of the drives. The quantum information of each qubit is encoded in the states $\left\vert g\right\rangle $ and $\left\vert f\right\rangle $, with the auxiliary state $\left\vert e\right\rangle $ used for realizing the controlled-NOT gate. The transitions $\left\vert g_{2}\right\rangle \longleftrightarrow \left\vert e_{2}\right\rangle $ and $\left\vert e_{2}\right\rangle \longleftrightarrow \left\vert f_{2}\right\rangle $ of Q$_{2}$ are driven by classical fields of angular frequencies $\left( \omega
		_{r}-\lambda _{2}\right) $ and $\left( \omega _{f,2}-\omega _{r}+\lambda
		_{2}\right) $, respectively. Here $\omega _{r}$ is the angular frequency of the
		resonator that is strongly coupled to $\left\vert g_{2}\right\rangle
		\longleftrightarrow \left\vert e_{2}\right\rangle $ with the coupling strength $\lambda _{2}$, and $\hbar \omega _{f,2}$ is the energy spacing between $%
		\left\vert f_{2}\right\rangle $ and $\left\vert g_{2}\right\rangle $. (b)
		Dressed states and energy levels with Q$_{1}$ initially in $\left\vert
		f_{1}\right\rangle $. When being initially in $\left\vert f_{1}\right\rangle$, Q$_{1}$ is effectively decoupled from the resonator due to the large detuning. The strong coupling between Q$_{2}$ and the resonator results in dressed states $\left\vert \psi _{n}^{\pm }\right\rangle $, whose energy levels are nonlinearly dependent on the coupling strength. The two driving fields are on resonance with the transitions $\left\vert g_{2}0\right\rangle \longleftrightarrow \left\vert \psi _{1}^{-}\right\rangle $
		and $\left\vert \psi _{1}^{-}\right\rangle \longleftrightarrow \left\vert
		f_{2}0\right\rangle $, respectively, but highly detuned from other
		transitions. (c) Dressed states and energy levels with Q$_{1}$ initially in $\left\vert g_{1}\right\rangle $. If initially in $\left\vert
		g_{1}\right\rangle $, Q$_{1}$, together with Q$_{2}$, is strongly coupled to the resonator, resulting in three dressed states $\left\vert \Phi _{1}^{\pm
		}\right\rangle $ and $\left\vert \Phi _{1}^{0}\right\rangle $ in the
		single-excitation subspace. The driving fields are highly detuned from
		transitions of $\left\vert g_{1}g_{2}0\right\rangle $ and $\left\vert
		g_{1}f_{2}0\right\rangle $ to these dressed states.}
	\label{f1}
\end{figure}

The system under consideration is composed of two qutrits coupled to a
resonator. Each qutrit has three basis states, as shown in Fig. \ref{f1}a, with $%
\left\vert g\right\rangle $ and $\left\vert f\right\rangle $ serving as two
logic states of a qubit, and $\left\vert e\right\rangle $, lying between $%
\left\vert g\right\rangle $ and $\left\vert f\right\rangle $, used as an
auxiliary state for realizing the controlled logic operation. For
simplicity, we will refer to the qutrits as qubits. As will be shown, the
control qubit (Q$_{1}$) remains in its computational space, while the target qubit (Q$_{2}$) has a probability of being
populated in the auxiliary level $\left\vert e\right\rangle $ during the
gate operation. The transition $\left\vert g\right\rangle
\longleftrightarrow \left\vert e\right\rangle $ of each qubit resonantly
interacts with the resonator, while $\left\vert f\right\rangle $ state is
effectively decoupled from the resonator. In the interaction picture, the
Hamiltonian describing the qubit-resonator interaction is given by 
\begin{equation}
	H_{\text{int}}= \hbar \sum_{j=1}^{2}\lambda _{j}\left( a\left\vert
	e_{j}\right\rangle \left\langle g_{j}\right\vert +a^{\dagger }\left\vert
	g_{j}\right\rangle \left\langle e_{j}\right\vert \right) ,
\end{equation}
where $a$ and $a^{\dagger }$ are the photonic annihilation and creation
operators for the resonator, $\lambda _{j}$ is the coupling strength between
the $j$th qubit and the resonator with angular frequency $\omega _{r}$. We here have set the energy of the ground state $\left\vert g\right\rangle $ for each qubit to be $0$. To realize the CNOT gate, the transition $\left\vert
g_{2}\right\rangle \longleftrightarrow \left\vert e_{2}\right\rangle $ of 
Q$_{2}$ is driven by a classical field with angular frequency $\left( \omega
_{r}-\lambda _{2}\right) $, and $\left\vert e_{2}\right\rangle
\longleftrightarrow \left\vert f_{2}\right\rangle $ is driven by a classical
field with angular frequency $\left( \omega _{f,2}-\omega _{r}+\lambda _{2}\right) 
$, where $\hbar \omega _{f,2}$ is the energy of Q$_{2}$'s state $\left\vert
f_{2}\right\rangle $ (Fig. 1a). The interaction between the second qubit and
the driving fields is described by%
\begin{equation}
	H_{\text{dr}}=\hbar \left[ \Omega _{ge}e^{i\lambda _{2}t}\left\vert
	e_{2}\right\rangle \left\langle g_{2}\right\vert -\Omega _{ef}e^{-i\lambda
		_{2}t}\left\vert f_{2}\right\rangle \left\langle e_{2}\right\vert \right]
	+h.c.
\end{equation}
where $\Omega _{ge}$ and $\Omega _{ef}$ denote the Rabi frequencies of the
two fields driving $\left\vert g_{2}\right\rangle \longleftrightarrow
\left\vert e_{2}\right\rangle $ and $\left\vert e_{2}\right\rangle
\longleftrightarrow \left\vert f_{2}\right\rangle $, respectively. We here
have assumed that the phases of the fields driving the transitions $%
\left\vert g\right\rangle \longleftrightarrow \left\vert e\right\rangle $
and $\left\vert e\right\rangle \longleftrightarrow \left\vert f\right\rangle 
$ are $0$ and $\pi $, respectively.

\begin{figure}%[htbp] 
	\centering
	%\fbox{\includegraphics[width=\linewidth]{Energy_Level.pdf}}
	\includegraphics[width=\linewidth]{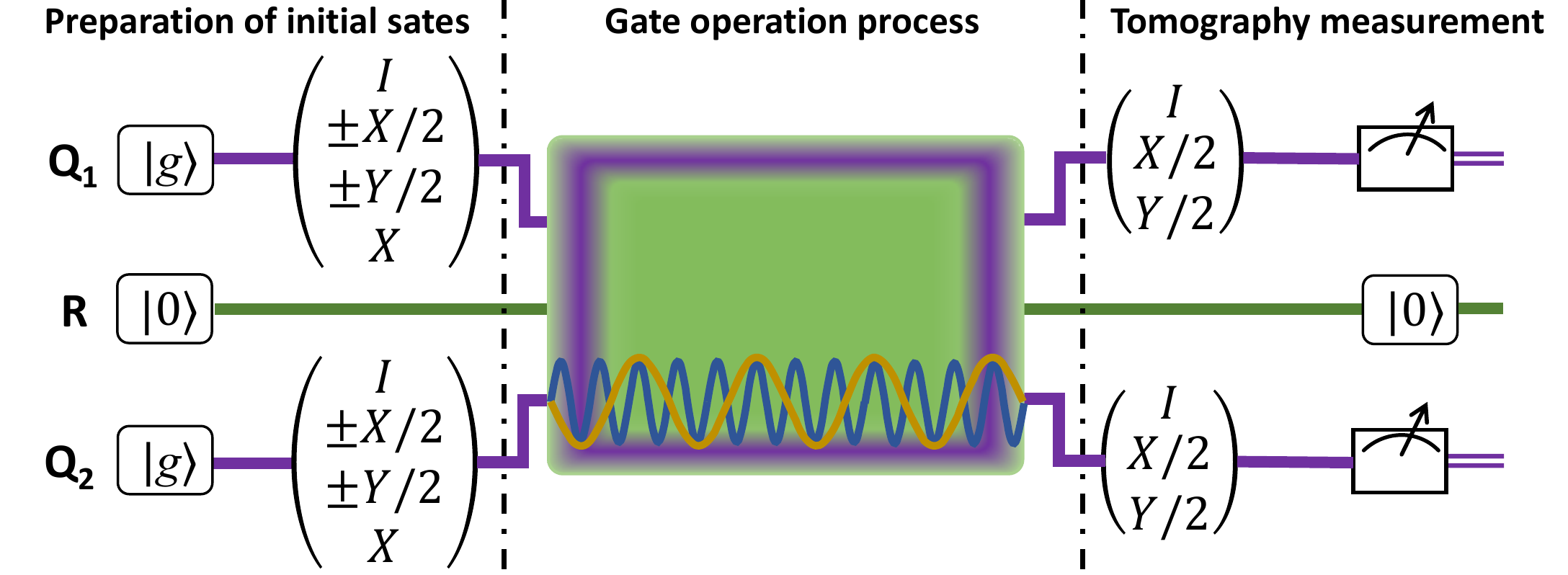}
	\caption{Pulse sequence. Before the gate operation,
		both qubits are initialized to their ground state at the corresponding idle
		frequencies, where single-qubit rotations are performed to prepare them in a
		product state. Then a Z pulse is applied to Q$_{1}$, tuning $\left\vert
		g_{1}\right\rangle \longleftrightarrow \left\vert e_{1}\right\rangle $ close to
		the resonator's frequency; Q$_{2}$ is subjected to a Z pulse, which brings $%
		\left\vert g_{2}\right\rangle \longleftrightarrow \left\vert e_{2}\right\rangle $
		to the resonator's frequency, and a driving pulse involving two frequency
		components respectively on resonance with the transitions $\left\vert
		g_{2}0\right\rangle \longleftrightarrow \left\vert \psi _{1}^{-}\right\rangle $
		and $\left\vert \psi _{1}^{-}\right\rangle \longleftrightarrow \left\vert
		f_{2}0\right\rangle $. After the CNOT gate, realized with these
		pulses, both qubits are tuned back to their idle frequencies for quantum
		state tomography.}
	\label{f2}
\end{figure}

\begin{figure}[htbp] 
	\centering
	\includegraphics[width=\linewidth]{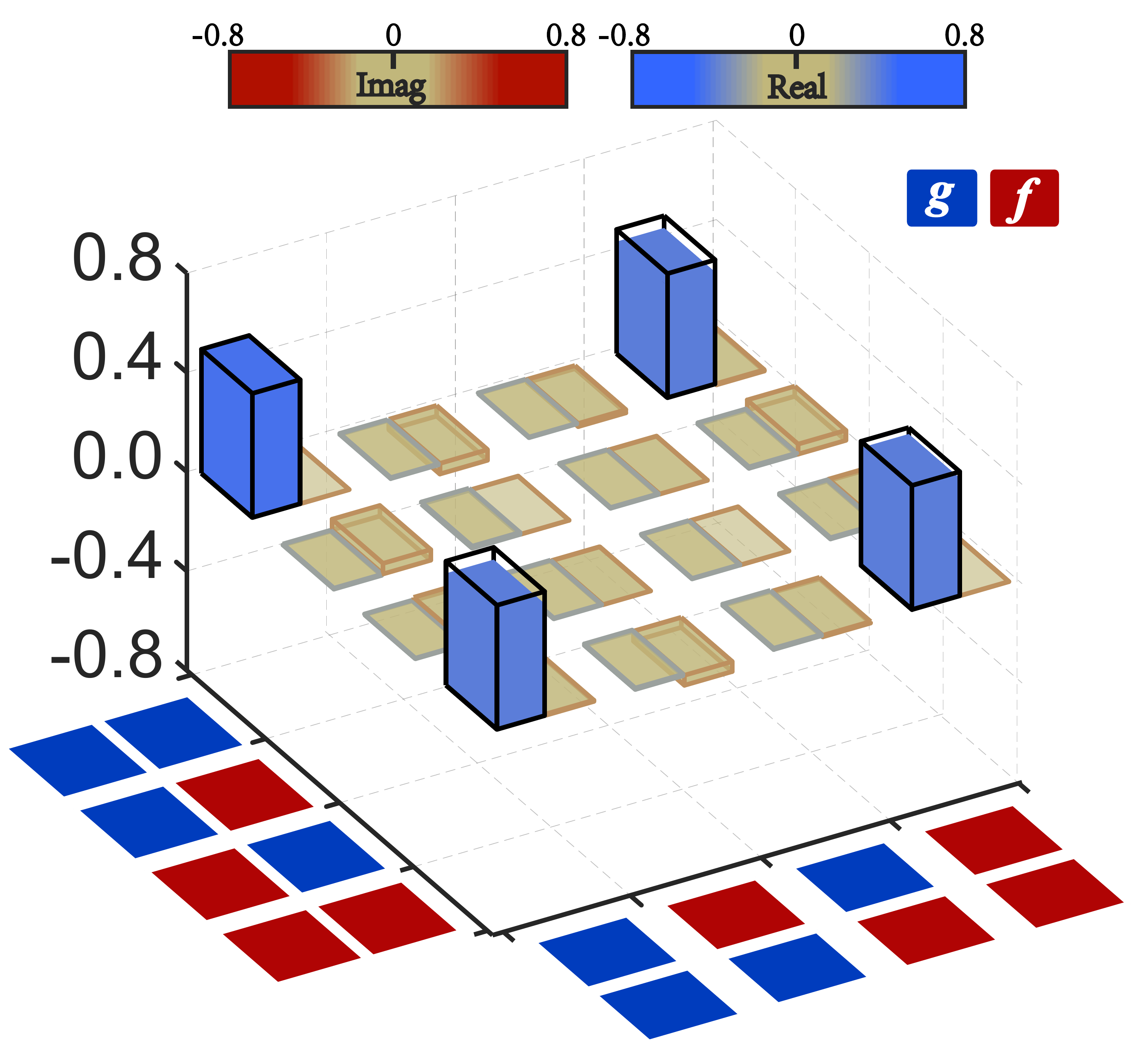}
	\caption{Measured density
		matrix of the output state with the input state $\left( \left\vert
		g_{1}g_{2}\right\rangle +\left\vert f_{1}g_{2}\right\rangle \right) /\sqrt{2}$. Each matrix element is characterized by two
		colorbars, one for the real part and the other for the imaginary part. The
		black wire frames denote the matrix elements of the ideal output states.}
	\label{f3}
\end{figure}

The strong couplings between the qubits and the resonator produce dressed
states, whose energy levels depend on the total excitation number as well as
on the number of qubits being initially populated in $\left\vert
g\right\rangle $. When the control qubit is in the state $\left\vert
f_{1}\right\rangle $, it does not interact with the resonator, and the
coupling between the target qubit and the resonator is described by the
Jaynes-Cummings model, whose eigenstates are given by 
\begin{equation}
	\left\vert \psi _{0}\right\rangle  =\left\vert g_{2}0\right\rangle,
\end{equation}
\begin{equation}
	\left\vert \psi _{n}^{\pm }\right\rangle  =\frac{1}{\sqrt{2}}\left(
	\left\vert e_{2}(n-1)\right\rangle \pm \left\vert g_{2}n\right\rangle
	\right) , n\geq 1.
\end{equation}
Here the second symbol in each ket denotes the photon number in the
resonator. The eigenenergies of the dressed states $\left\vert \psi
_{n}^{\pm }\right\rangle $ are $\hbar \left( n\omega _{r}\pm \sqrt{n}\lambda
_{2}\right) $. We here consider the case that the resonator is initially in
the vacuum state $\left\vert 0\right\rangle $. Consequently, the classical
fields resonantly couple the states $\left\vert g_{2}0\right\rangle $ and $%
\left\vert f_{2}0\right\rangle $ to the single-excitation dressed state $%
\left\vert \psi _{1}^{-}\right\rangle $, respectively, as sketched in Fig.
\ref{f1}b. We suppose that $\Omega _{ge}$ and $\Omega _{ef}$ are much smaller than $%
\lambda _{2}$, so that the classical fields cannot drive the transitions
from $\left\vert \psi _{1}^{-}\right\rangle $ to $\left\vert \psi _{2}^{\pm }\right\rangle $ due to the large detunings.
However, these
off-resonant couplings shift the energy levels of $\left\vert \psi
_{1}^{- }\right\rangle $ by $-2 \hbar \delta _{1}$, with $\delta
_{1}=2\Omega _{ge}^{2}/\lambda _{2}$ (see Supplemental Material). Furthermore, off-resonant coupling to
$\left\vert h_{1}\right\rangle \left\vert g_{2}0\right\rangle $ and $%
\left\vert e_{1}\right\rangle \left\vert \psi _{2}^{\pm }\right\rangle $
shifts the energy level of $\left\vert f_{1}\right\rangle \left\vert \psi
_{1}^{\pm }\right\rangle $ by an amount of $-\hbar\delta _{2}$, where $\delta_2=9\lambda
_{1}^{2}/4\alpha _{1}$ (see Supplemental Material), and $\left\vert h_{1}\right\rangle $ is the fourth level of Q$_{1}$ and $\alpha _{1}$ is its anharmonicity ($\alpha_j=2\omega_{e,j}-\omega_{f,j}$, $j=1,2$). To
compensate for these shifts, the angular frequency of the field driving $\left\vert
g_{2}\right\rangle \longleftrightarrow \left\vert e_{2}\right\rangle $
should be set to $\omega _{d,1}=\omega _{r}-\lambda _{2}-\delta _{1}-\delta
_{2}$, while that of the field driving $\left\vert
e_{2}\right\rangle \longleftrightarrow \left\vert f_{2}\right\rangle $
should be set to $\omega _{d,2}=\omega _{f,2}-\omega _{r}+\lambda
_{2}+\delta _{1}+\delta _{2}$. 
With this setting and performing the transformation $\exp \left( iH_{\text{%
		int}}t/\hbar \right) $, the system dynamics associated with Q$_{1}$'s state $%
\left\vert f_{1}\right\rangle $ can be described by the effective Hamiltonian%
\begin{equation}\label{e5}
	H_{\text{eff}}=\hbar\Omega \left[ \cos \frac{\phi }{2}\left\vert
	g_{2}0\right\rangle \left\langle \psi _{1}^{-}\right\vert +\sin \frac{\phi }{%
		2}\left\vert f_{2}0\right\rangle \left\langle \psi _{1}^{-}\right\vert %
	\right] \left\vert f_{1}\right\rangle \left\langle f_{1}\right\vert +h.c.,
\end{equation}
where 
\begin{equation}
	\Omega  =\sqrt{\Omega _{ge}^{2}+\Omega _{ef}^{2}}/\sqrt{2}, 
\end{equation}	
\begin{equation}
	\tan \frac{\phi }{2} =\Omega _{ef}/\Omega _{ge}.
\end{equation}

When Q$_{1}$ is initially in the state $\left\vert g_{1}\right\rangle $, it
is also strongly coupled to the resonator, and there are three dressed
states in the single-excitation subspace: 
\begin{equation}
	\left\vert \Phi _{1}^{0}\right\rangle =\left( -\sin \theta \left\vert
	e_{1}g_{2}0\right\rangle +\cos \theta \left\vert g_{1}e_{2}0\right\rangle
	\right),
\end{equation}	
\begin{equation}
	\left\vert \Phi _{1}^{\pm }\right\rangle =\frac{1}{\sqrt{2}}\left[ \left(
	\cos \theta \left\vert e_{1}g_{2}0\right\rangle +\sin \theta \left\vert
	g_{1}e_{2}0\right\rangle \right) \pm \left\vert
	g_{1}g_{2}1\right\rangle \right] ,
\end{equation}%
where $\tan \theta =\lambda _{2}/\lambda _{1}$. The corresponding
eigenenergies are $\hbar \omega _{e}$ and $\hbar \left( \omega _{e}\pm \sqrt{%
	\lambda _{1}^{2}+\lambda _{2}^{2}}\right) $, as shown in Fig. \ref{f1}c. When $%
\left( \sqrt{\lambda _{1}^{2}+\lambda _{2}^{2}}-\lambda _{2}\right) $ is
much larger than $\Omega _{ge}$ and $\Omega _{ef}$, the qubits cannot make
any transition between each of these single-excitation dressed states and
the state $\left\vert g_{1}g_{2}0\right\rangle $ or $\left\vert
g_{1}f_{2}0\right\rangle $ as each of these transitions is highly detuned
from the driving fields. As a consequence, Q$_{2}$ is not affected by the
driving fields when Q$_{1}$ is initially in the state $\left\vert
g_{1}\right\rangle $. Therefore, the system dynamics is described by the
effective Hamiltonian of Eq. (\ref{e5}). The evolution of the initial basis states $%
\left\vert c_{1}d_{2}0\right\rangle $ ($c,d=g,f$) are given by 
\begin{equation}\label{e10}
	\left\vert \psi _{cd}(t)\right\rangle =\exp \left( -i\int_{0}^{t}H_{\text{eff%
	}}dt/\hbar \right) \left\vert c_{1}d_{2}0\right\rangle . 
\end{equation}
When $\Omega _{ef}/\Omega _{ge}$ remains unchanged during the interaction,
the evolution satisfies the parallel-transport condition $\left\langle \psi
_{cd}(t)\right\vert H_{\text{eff}}\left\vert \psi _{c^{^{\prime
	}}d^{^{\prime }}}(t)\right\rangle =0$, and hence is purely geometric. If the
Rabi frequencies of the driving fields and the interaction time are
appropriately chosen so that $\int_{0}^{T}\Omega dt=\pi $, the degenerate
qubit subspace undergoes a cyclic evolution. Consequently, the qubits return
to the computational space $\left\{ \left\vert g_{1}g_{2}\right\rangle
,\left\vert g_{1}f_{2}\right\rangle ,\left\vert f_{1}g_{2}\right\rangle
,\left\vert f_{1}f_{2}\right\rangle \right\} $ with the resonator left in
the vacuum state $\left\vert 0\right\rangle $ after the time $T$. With this
setting, the evolution operator of the qubits in the computational basis is 
\begin{eqnarray}
	U=\left( 
	\begin{array}{cccc}
		1 & 0 & 0 & 0 \\ 
		0 & 1 & 0 & 0 \\ 
		0 & 0 & -\cos \phi & \sin \phi \\ 
		0 & 0 & \sin \phi & \cos \phi%
	\end{array}%
	\right) , 
\end{eqnarray}
which is a non-Abelian holonomy. For $\phi =\pi /2$, i.e., $\Omega
_{ge}=\Omega _{ef}$, this corresponds to a CNOT gate, which flips the state
of the target qubit conditional on the control qubit being in the state $%
\left\vert f_{1}\right\rangle $. 
\iffalse
\textcolor{red}{We note that during the gate operation, the
two qubits have a probability of being populated in $\left\vert
f_{1}e_{2}\right\rangle $, which is significantly coupled to $\left\vert
e_{1}f_{2}\right\rangle $ via virtual photon exchange when the anharmonicity
of $Q_{1}$ is close to that of $Q_{2}$. To suppress this coupling, $Q_{1}$
should be detuned from $Q_{2}$ by an amount much larger than $\lambda
_{1}\lambda _{2}/\alpha $. This detuning slightly changes the energy level
configuration of the dressed states associated with $Q_{1}$'s initial state $%
\left\vert g_{1}\right\rangle $, but does not affect the gate dynamics.}
\fi

\begin{figure}[htbp] 
	\centering
	\includegraphics[width=\linewidth]{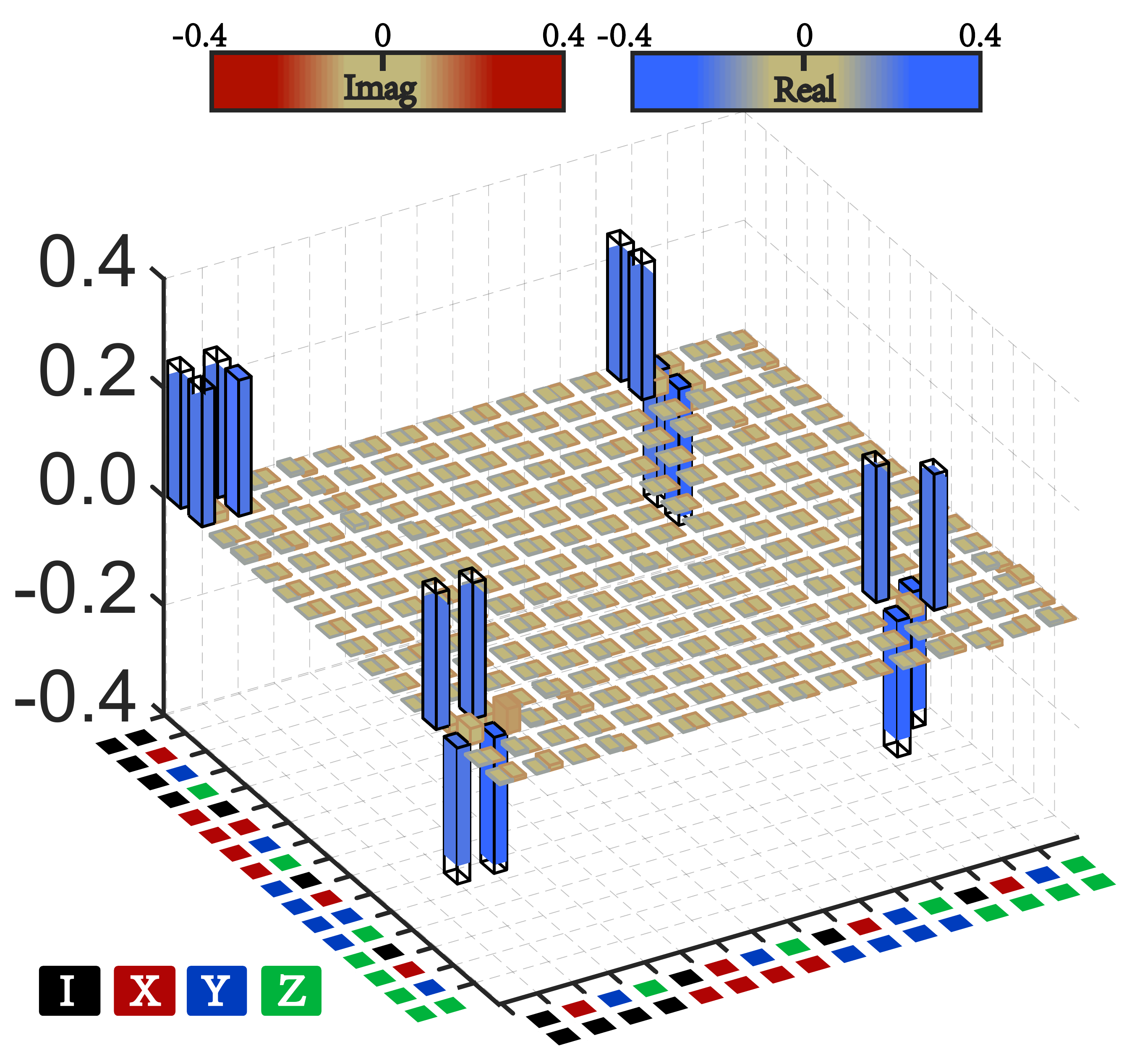}
	\caption{ Measured process matrix for the realized CNOT gate.
		The process matrix is measured by preparing a set of 36 distinct input
		product states in the computational basis $\left\{ \left\vert
		g_{1}g_{2}\right\rangle ,\left\vert g_{1}f_{2}\right\rangle ,\left\vert
		f_{1}g_{2}\right\rangle ,\left\vert f_{1}f_{2}\right\rangle \right\} $ and
		reconstructing the density matrices for these states and for the output
		states produced by the CNOT gate. The $\left\vert e\right\rangle $%
		-state populations of Q$_{1}$ and Q$_{2}$, averaged over the 36 output states, are
		2.2\% and 2.8\%, respectively.}
	\label{f4}
\end{figure}

\section{Experimental implementation}
The experiment is performed in a superconducting circuit involving five
frequency-tunable qubits, labeled from Q$_{1}$ to Q$_{5}$, coupled to a
resonator with a fixed frequency $\omega _{r}/2\pi=$5.584 GHz \cite{Song_NatCommun2017,Ning_PRL2019,Yang_npj_Quantum Information2021}. In our
experiment, Q$_{1}$ and Q$_{2}$, whose anharmonicities are $2\pi\times 242$ MHz and  $2\pi\times 249$ MHz,
are used as the control and target qubits, respectively. The on-resonance
coupling strengths of the $g$-$e$ transitions of Q$_{1}$ and Q$_{2}$ to the
resonator are respectively $\lambda _{1}$=$2\pi\times 20.8$ MHz and $\lambda _{2}$=$2\pi\times 19.9$ MHz.
The energy relaxation time T$_{1}$ and pure Gaussian dephasing time T$%
_{2}^{\ast }$ for the basis state $\left\vert f\right\rangle $ of Q$_{1}$ (Q$%
_{2}$) are 13.0 (10.7) $\mu $s and 2.1 (1.5) $\mu $s, while those for the intermediate
state $\left\vert e\right\rangle $ are 23.9 (15.9) $\mu $s and 2.7 (2.1) $\mu $s,
respectively. The other qubits are on far off-resonance with the resonator
so that their interactions with the resonator are effectively switched off
throughout the gate operation. We note that during the gate operation, the
two qubits have a probability of being populated in $\left\vert
f_{1}e_{2}\right\rangle $, which is significantly coupled to $\left\vert
e_{1}f_{2}\right\rangle $ via virtual photon exchange as the two qubits
almost have the same anharmonicity $\alpha \simeq 2\pi\times240$ MHz. To suppress this
coupling, Q$_{1}$ should be detuned from Q$_{2}$ by an amount much larger than $%
\lambda _{1}\lambda _{2}/\alpha $. This detuning slightly changes the energy
level configuration of the dressed states associated with Q$_{1}$'s initial state 
%TCIMACRO{\TEXTsymbol{\vert}}%
%BeginExpansion
%\mbox{$\vert$}%
%EndExpansion
$\left\vert g_1\right\rangle$, but does not affect the gate dynamics.

As shown in Fig. \ref{f2}, the experiment starts with the initialization of Q$_{1}$ and Q$_{2}$ to the ground state $\left\vert g\right\rangle $ at their
idle frequencies $5.47$ GHz and $5.43$ GHz, respectively, which is followed by the
preparation of each qubit in one of the six states $\{\left\vert
g\right\rangle$, $\left( \left\vert g\right\rangle -i\left\vert f\right\rangle
\right) /\sqrt{2}$, $\left( \left\vert g\right\rangle +i\left\vert f\right\rangle
	\right) /\sqrt{2}$, $\left( \left\vert g\right\rangle +\left\vert
f\right\rangle \right) /\sqrt{2}$, $\left( \left\vert g\right\rangle -\left\vert
	f\right\rangle \right) /\sqrt{2}$, $\left\vert f\right\rangle$$\}$.
Except $\left\vert g\right\rangle $, each of the other single-qubit states
is produced by a $g$-$e$ $\pi /2$- or $\pi $-pulse followed by a $e$-$f$ $%
\pi $-pulse. 
\iffalse
After these effective single-qubit rotations, these two qubits
are prepared in a product state. We then apply square Z pulses to both
qubits, tuning their $g$-$e$ transition frequencies to \textcolor{red}{$5.58$} GHz and \textcolor{red}{$5.584$} GHz and
thus switching on their interactions with the resonator.
\fi 
After these effective single-qubit rotations, these two qubits are prepared in a product state. We then apply square Z pulses to both qubits, tuning their $%
\left|g \right\rangle\longleftrightarrow \left|e\right\rangle$ transition frequencies to $5.58$ GHz and $5.584$ GHz and
thus switching on their interactions with the resonator. Accompanying these
Z pulses, a driving pulse composed of two components with frequencies of
$5.565$ GHz and $5.369$ GHz is applied to Q$_{2}$, resonantly connecting the
computational states $\left\vert g_{2}0\right\rangle \ $and $\left\vert
f_{2}0\right\rangle $ to the dressed state $\left\vert \psi
_{1}^{-}\right\rangle $. The Rabi frequencies of these driving fields are $\Omega_{ge}=\Omega_{ef}=2\pi\times2.2$ MHz.
Since the resonator is initially in the vacuum
state, the system dynamics is governed by the effective Hamiltonian (\ref{e5}) and
the time evolution given by Eq. (\ref{e10}). After a duration of $\tau =205$ ns, the
CNOT gate is realized.

One of the most important features of the CNOT gate is that it can convert a
two-qubit product state into an entangled state. In particular, when the
control qubit is initially in the superposition state $\left( \left\vert
g_{1}\right\rangle +\left\vert f_{1}\right\rangle \right) /\sqrt{2}$ and the
target state in $\left\vert g_{2}\right\rangle $, they will evolve to the
maximally entangled state $\left( \left\vert g_{1}g_{2}\right\rangle +\left\vert f_{1}f_{2}\right\rangle \right) /\sqrt{2}$ after this gate. We measure this
output state by quantum state tomography. This is realized by subsequently
biasing each of the two qubits back to its idle frequency right after the
gate operation, applying an $e$-$f$ $\pi$-pulse to each qubit, and measuring its state along one of the three
orthogonal (X, Y, and Z) axes of the corresponding Bloch sphere with respect
to the basis $\left\{ \left\vert g\right\rangle ,\left\vert e\right\rangle
	 \right\}$. The Z measurement is directly realized by state readout, while
the X (Y) measurement realized by the combination of a $g$-$e$ $\pi /2$-pulse and state readout. The reconstructed output
two-qubit density matrix is displayed in Fig. \ref{f3}, which has a fidelity of
$0.935\pm0.016$ to the ideal maximally entangled state, and a concurrence of $0.888\pm0.029$.

To fully characterize the performance of the implemented CNOT gate, we
prepare a full set of 36 distinct two-qubit input states before the
two-qubit gates, and measure these states and the corresponding output
states. With these measured results, the process matrix for the gate
operation is reconstructed. The measured process matrix, $\chi _{\text{meas}%
} $, is presented in Fig. \ref{f4}. The gate fidelity, defined as $F=tr\left( \chi
_{\text{id}}\chi _{\text{meas}}\right) $, is $0.905\pm0.008$, where $\chi _{\text{id}}$ is
the ideal process matrix. The measured fidelity is in well agreement with
the numerical simulation based on the Lindblad master equation, which yields
a fidelity of 0.908. One of the error sources is the transitions from $\left\vert
g_{1}g_{2}0\right\rangle $ and $\left\vert g_{1}f_{2}0\right\rangle $ to 
$\left\vert \Phi _{1}^{0}\right\rangle $ and $\left\vert \Phi _{1}^{\pm
}\right\rangle $ and the transition from $\left\vert \psi
_{1}^{-}\right\rangle $ to $\left\vert \psi _{2}^{-}\right\rangle $ induced
by the drive, which cause quantum information leakage to the
noncomputational space. Such a leakage error can be mitigated
through the improvement of the qubit's nonlinearity or by balancing the
drive amplitude and the gate operation time provided the qubits' coherence
is bettered, which allows the gate fidelity to be increased by about 6.5\%
(see Supplemental Material). On the other hand, the qubits' energy
relaxation and their dephasings contribute about 1.8\% and 1.6\% of the
error, respectively. Our further numerical simulations show that the CNOT
gate with a fidelity above 99\% can be obtained with sufficiently large
qubit's nonlinearity $\alpha _{j}$ and qubit-resonator coupling strength $%
\lambda _{j}$. For instance, with the parameters $\lambda _{j}/2\pi=110$ MHz, $%
\alpha _{j}/2\pi$ $=-3.69$ GHz \cite{Yurtalan_arXiv2020,You_PRB2007}, $\Omega _{ge}/2\pi=\Omega _{ef}/2\pi=5.9$ MHz, $%
T_{1}=60$ $\mu $s, and $T_{2,j}^{\ast }$ $=86$ $\mu $s, we find a CNOT gate
with the operation time about 87 ns and the fidelity of 0.991, which is at
the surface code threshold for fault tolerance \cite{Fowler_PRA2012,Barends_Nature2014,Barends_Nature2015}. We note this gate
is robust against the frequency fluctuations of the driving fields. Suppose that 
the angular frequencies of these drives deviate from the desired values by an amount of $\delta\omega=2\pi\times100$ kHz. The infidelity incurred by this deviation is 
about $[\pi(\delta\omega)^2/(8\Omega_{ge/ef}^2)]^2\simeq 0.1\% $.

\section{Conclusion}

In conclusion, we have proposed and demonstrated a scheme for implementing a
non-Abelian geometric gate between two superconducting qubits, whose ground
and second excited states act as the computational basis states. The
conditional dynamics is realized by resonantly driving the transitions
between the basis states of the target qubit to the single-excitation
dressed states formed by this qubit and the resonator. This entangling gate,
together with the previously demonstrated non-Abelian geometric single-qubit
gates \cite{Abdumalikov_Nature2013,Xu_PRL2018,Yan_PRL2019,Zhang_arXiv1811}, constitutes a universal set of holonomic gates for realizing
quantum computation with superconducting qubits. The method can be directly
applied to other systems composed of qubits coupled to a bosonic mode,
including cavity QED and ion traps.

%This work was supported by the National Natural Science Foundation of China
%(Grant No. 11874114, No. 11674060, and No. 11875108), and the Strategic
%Priority Research Program of Chinese Academy of Sciences (Grant No.
%XDB28000000).

\medskip

\noindent\textbf{Funding.} National Natural Science Foundation of China (Grant No. 11674060, No. 11874114, No. 11875108, No. 11934018, No. 11904393, and No. 92065114),
Strategic Priority Research Program of Chinese Academy of Sciences (Grant No. XDB28000000), and Beijing Natural Science Foundation (Grant No. Z200009).
\medskip

\noindent\textbf{Acknowledgment.} We thank Haohua Wang at Zhejiang University for technical support.

\medskip

\noindent\textbf{Disclosures.} The authors declare no conflicts of interest.

\medskip

\noindent\textbf{Contributions.} S.-B.Z. conceived the experiment. K.X., W. N., and Z.-B.Y. performed the experiment and analyzed the data with the assistance of X.-J.H. and P.-R.H. H.L. and D.Z. provided the devices used for the experiment. S.-B.Z., Z.-B.Y., K.X., and H.F. wrote the manuscript with feedbacks from all authors. The experiment was performed at Fuzhou University.

%\section {xx}
%\label{details tran T2}

%\begin{thebibliography}{99}

%\bibitem{} A. Einstein, B. Podolsky, and N. Rosen, Phys. Rev. 47, 777 (1935).

%


\begin{thebibliography}{1}
	\newcommand{\enquote}[1]{``#1''}
	\bibitem{Berry_PRSL1984} M. V. Berry,
	\enquote{Quantal phase-factors accompanying adiabatic changes,}
	Proc. R. Soc. Lond. A {\bf 392}, 45 (1984). 
	%\bibitem{Berry_PRSL1984} M. V. Berry, "Quantal phase-factors accompanying adiabatic
	%changes", Proc. R. Soc. Lond. A 392, 45 (1984). 
	
	\bibitem{Aharonov_PRL1987} Y. Aharonov and A. Anandan, 
	\enquote{Phase change during a cyclic quantum evolution,} 
	Phys. Rev. Lett. {\bf 58}, 1593 (1987). 
	%\bibitem{Aharonov_PRL1987} Y. Aharonov, and A. Anandan, Phase change during a cyclic quantum
	%evolution. Phys. Rev. Lett. 58, 1593-1596 (1987).[2] 
	
	\bibitem{Wilczek_PRL1984} F. Wilczek and A. Zee, 
	\enquote{Appearance of gauge structure in simple dynamical systems,}
	Phys. Rev. Lett. {\bf 52}, 2111 (1984). 
	%\bibitem{Wilczek_PRL1984} Wilczek, F. \& Zee, A. Appearance of gauge structure in simple
	%dynamical systems. Phys. Rev. Lett. 52, 2111-2114 (1984).[3] 
	
	\bibitem{Chiara_PRL2003} G. D. Chiara and G. M. Palma, 
	\enquote{Berry phase for a spin 1/2 particle in a classical fluctuating field,}
	Phys. Rev. Lett. {\bf 91}, 090404 (2003). 
	%\bibitem{Chiara_PRL2003} Chiara, G. D. \& Palma, G. M. Berry phase for a spin 1/2 particle
	%in a classical fluctuating field. Phys. Rev. Lett. 91, 090404 (2003).[4] 
	
	\bibitem{Filipp_PRL2009} S. Filipp, J. Klepp, Y. Hasegawa, C. Plonka-Spehr, U. Schmidt, P. Geltenbort, and H. Rauch, \enquote{Experimental demonstration of the stability of Berry's phase for a spin-1/2 particle,}
	Phys. Rev. Lett. {\bf 102}, 030404 (2009). 
	%\bibitem{Filipp_PRL2009} Filipp, S. et al. Experimental demonstration of the stability of
	%Berry's phase for a spin-1/2 particle. Phys. Rev. Lett. 102, 030404 (2009). [5] 
	
	\bibitem{Carollo_PRL2003} A. Carollo, I. Fuentes-Guridi, M. F. Santos, and V. Vedral,           
	\enquote{Geometric phase in open systems,} 
	Phys. Rev. Lett. {\bf 90}, 160402 (2003). 
	%\bibitem{Carollo_PRL2003} Carollo, A., Fuentes-Guridi, I., Santos, M. F. \& Vedral, V.
	%Geometric phase in open systems. Phys. Rev. Lett. 90, 160402 (2003).[6] 
	
	\bibitem{Zheng_PRA2015} S.-B. Zheng, 
	\enquote{Geometric phase for a driven quantum field subject to decoherence,} 
	Phys. Rev. A {\bf 91}, 052117 (2015). 
	%\bibitem{Zheng_PRA2015} Zheng, S. B. Geometric phase for a driven quantum field subject
	%to decoherence. Phys. Rev. A 91, 052117 (2015).[7] 
	
	\bibitem{Zanardi_PLA1999} P. Zanardi and M. Rasetti,
	\enquote{Holonomic quantum computation,} 
	Phys. Lett. A {\bf 264}, 94 (1999). 
	%\bibitem{Zanardi_PLA1999} Zanardi, P. \& Rasetti, M. Holonomic quantum computation. Phys.
	%Lett. A 264, 94-99 (1999).[8] 
	
	\bibitem{Pachos_PRA2000} J. Pachos, P. Zanardi, and M. Rasetti,
	\enquote{Non-Abelian Berry connections for quantum computation,}  
	Phys. Rev. A {\bf 61}, 010305(R) (2000).
	%\bibitem{Pachos_PRA2000} Pachos, J., Zanardi, P. \& Rasetti, M. Non-Abelian Berry
	%connections for quantum computation. Phys. Rev. A 61,010305(R) (2000).[9] 
	
	\bibitem{Jones_Nature2000} J. A. Jones, V. Vedral, A. Ekert, and G. Castagnoli,
	\enquote{Geometric quantum computation with NMR,} 
	Nature (London) {\bf 403}, 869 (2000). 
	%\bibitem{Jones_Nature2000} Jones, J. A., Vedral, V., Ekert, A. \& Castagnoli, G. Geometric
	%quantum computation with NMR. Nature 403, 869-871 (2000). [10] 
	
	\bibitem{Wang_PRL2001} X.-B. Wang and M. Keiji,
	\enquote{Nonadiabatic conditional geometric phase shift with NMR,}
	Phys. Rev. Lett. {\bf 87}, 097901 (2001). 
	%\bibitem{Wang_PRL2001} X. B. Wang and M. Keiji, Nonadiabatic Conditional Geometric Phase
	%Shift with NMR, Phys. Rev. Lett. 87, 097901 (2001).[11] 
	
	\bibitem{Zhu_PRL2002} S.-L. Zhu and Z.-D. Wang,
	\enquote{Implementation of universal quantum gates based on nonadiabatic geometric phases,}
	Phys. Rev. Lett. {\bf 89}, 097902 (2002).
	%\bibitem{Zhu_PRL2002} S. L. Zhu and Z. D. Wang, Implementation of Universal Quantum
	%Gates Based on Nonadiabatic Geometric Phases, Phys. Rev. Lett. 89, 097902 
	%(2002). [12] 
	
	\bibitem{Nazir_PRA2002} A. Nazir, T. Spiller, and W. J. Munro,
	\enquote{Decoherence of geometric phase gates,} 
	Phys. Rev. A {\bf 65}, 042303 (2002). 
	%\bibitem{Nazir_PRA2002} A. Nazir, T. Spiller, and W. J. Munro, Decoherence of geometric
	%phase gates, Phys. Rev. A 65, 042303 (2002). [13] 
	
	\bibitem{Blais_PRA2003} A. Blais and A.-M. S. Tremblay, 
	\enquote{Effect of noise on geometric logic gates for quantum computation,} 
	Phys. Rev. A {\bf 67}, 012308 (2003).
	%\bibitem{Blais_PRA2003} A. Blais and A.-M. S. Tremblay, Effect of noise on geometric
	%logic gates for quantum computation, Phys. Rev. A 67, 012308 (2003). [14] 
	
	\bibitem{ZhangJ_PRA2018} J. Zhang, S. J. Devitt, J. Q. You, and F. Nori, 
	\enquote{Holonomic surface codes for fault-tolerant quantum computation,}
	Phys. Rev. A {\bf 97}, 022335 (2018).
	%%%%%%%%%%%%%%%%%%%%%%%%%%%%%%%%%%%%%%%%%%%%%%%%%%%%%%%%%
	\bibitem{Chen_arXiv2012} Y.-H. Chen, W. Qin, R. Stassi, X. Wang, and F. Nori,
	\enquote{Generation of Fock-State Superpositions and binomial-code holonomic gates
		via dressed intermediate states in the ultrastrong light-matter coupling
		Regime,} arXiv:2012.06090. 
	
	\bibitem{Leibfried_Nature2003} D. Leibfried, B. DeMarco, V. Meyer, D. Lucas, M. Barrett,
	J. Britton, W. M. Itano, B. Jelenkovi$\acute{c}$, C. Langer, T. Rosenband, and D. J. Wineland,
	\enquote{Experimental demonstration of a robust, high-fidelity geometric two ion-qubit phase gate,}
	Nature (London) {\bf 422}, 412 (2003).
	%\bibitem{Leibfried_Nature2003} Leibfried, D. et al. Experimental demonstration of a robust,
	%high-fidelity geometric two ion-qubit phase gate. Nature 422, 412-415 (2003). [17] 
	
	\bibitem{Benhelm_NaturePhysics2008} J. Benhelm, G. Kirchmair, C. F. Roos, and R. Blatt,
	\enquote{Towards fault-tolerant quantum computing with trapped ions,}
	Nat. Phys {\bf 4}, 463 (2008).
	%\bibitem{Benhelm_NaturePhysics2008} Benhelm, J., Kirchmair, G., Roos, C. F., \& Blatt, R. Towards
	%fault-tolerant quantum computing with trapped ions. Nature Physics 4,
	%463-466 (2008). [18] 
	
	\bibitem{Ballance_PRL2016} C. J. Ballance, T. P. Harty, N. M. Linke, M. A. Sepiol, and D. M.
	Lucas,
	\enquote{High-fidelity quantum logic gates using trapped-ion hyperfine qubits,}
	Phys. Rev. Lett. {\bf 117}, 060504 (2016).
	%\bibitem{Ballance_PRL2016} C. J. Ballance, T. P. Harty, N. M. Linke, M. A. Sepiol, and D. M.
	%Lucas, Phys. Rev. Lett. 117, 060504 (2016).[19] 
	
	\bibitem{Gaebler_PRL2016} J. P. Gaebler, T. R. Tan, Y. Lin, Y. Wan, R. Bowler, A. C. Keith,
	S. Glancy, K. Coakley, E. Knill, D. Leibfried, and D. J. Wineland,
	\enquote{High-fidelity universal gate set for $^9Be^+$ ion qubits,}
	Phys. Rev. Lett. {\bf 117}, 060505 (2016).
	%\bibitem{Gaebler_PRL2016} J. P. Gaebler, T. R. Tan, Y. Lin, Y. Wan, R. Bowler, A. C. Keith,
	%S. Glancy, K. Coakley, E. Knill, D. Leibfried, and D. J. Wineland, Phys.
	%Rev. Lett. 117, 060505 (2016). [20] 
	
	\bibitem{Song_NatCommun2017} C. Song, S.-B. Zheng, P. Zhang, K. Xu, L. Zhang, Q. Guo,
	W. Liu, D. Xu, H. Deng, K. Huang, D. Zheng, X. Zhu, H. Wang,
	\enquote{Continuous-variable geometric phase and its manipulation for quantum computation in a superconducting circuit,} Nat. Commun. {\bf 8}, 1061 (2017).
	%\bibitem{Song_NatCommun2017} C. Song et al., Nat. Commun. 8, 1061 (2017). [21] 
	
	\bibitem{Xu_PRL2020} Y. Xu, Z. Hua, T. Chen, X. Pan, X. Li, J. Han, W. Cai, Y. Ma, H. Wang, Y. P. Song, Z.-Y. Xue, and L. Sun,
	\enquote{Experimental implementation of universal nonadiabatic geometric quantum gates in a superconducting circuit,} Phys. Rev. Lett. {\bf 124}, 230503 (2020).
	%\bibitem{Xu_PRL2020} Yuan Xu, Ziyue Hua, Tao Chen, Xiaoxuan Pan, Xuegang Li, Jiaxiu
	%Han, Weizhou Cai, Yuwei Ma, Haiyan Wang, Yipu Song, Zheng-Yuan Xue, Luyan
	%Sun, Phys. Rev. Lett. 124, 230503 (2020). [22] 
	\bibitem{Xu_PRL2020_Gate} Y. Xu, Y. Ma, W. Cai, X. Mu, W. Dai, W. Wang, L. Hu, X.
	Li, J. Han, H. Wang, Y. P. Song, Z.-B. Yang, S.-B. Zheng,
	and L. Sun, \enquote{Demonstration of Controlled-Phase Gates
	Between Two Error-Correctable Photonic Qubits,} Phys.
	Rev. Lett. {\bf 124}, 120501 (2020).
	
	
	\bibitem{Sjoqvist_NJP2012} E. Sj\"{o}qvist, D. M. Tong, L. M. Andersson, B. Hessmo, M. Johansson, and K. Singh,
	\enquote{Nonadiabatic holonomic quantum computation,}
	N. J. Phys. {\bf 14}, 103035 (2012).
	%\bibitem{Sjoqvist_NJP2012} Sj\"{o}qvist, E et al. Nonadiabatic holonomic quantum
	%computation. N. J. Phys. 14, 103035 (2012). [23] 
	
	\bibitem{Johansson_PRA2012} M. Johansson, E. Sj\"{o}qvist, L. M. Andersson, M. Ericsson, B.
	Hessmo, K. Singh, and D. M. Tong,
	\enquote{Robustness of non-adiabatic holonomic gates,}
	Phys. Rev. A {\bf 86}, 062322 (2012).
	%\bibitem{Johansson_PRA2012} M. Johansson, E. Sj\"{o}qvist, L. M. Andersson, M. Ericsson, B.
	%Hessmo, K. Singh, and D. M. Tong, Robustness of non-adiabatic holonomic
	%gates, Phys. Rev. A 86, 062322 (2012). [24] 
	
	\bibitem{Zheng_PRA2016} S.-B. Zheng, C.-P. Yang, and F. Nori,
	\enquote{Comparison of the sensitivity to systematic errors between nonadiabatic non-Abelian geometric gates and their dynamical counterparts,}
	Phys. Rev. A {\bf 93}, 032313 (2016).
	%\bibitem{Zheng_PRA2016} Shi-Biao Zheng, Chui-Ping Yang, and Franco Nori, Phys. Rev. A 93,
	%032313 (2016). [25] 
	
	\bibitem{Feng_PRL2013} G. Feng, G. Xu, and G. Long,
	\enquote{Experimental realization of nonadiabatic holonomic quantum computation,}
	Phys. Rev. Lett. {\bf 110}, 190501 (2013).
	%\bibitem{Feng_PRL2013} Feng, G., Xu, G. \& Long, G. Experimental realization of
	%nonadiabatic holonomic quantum computation. Phys. Rev. Lett. 110, 190501
	%(2013). [26] 
	
	\bibitem{Zu_Nature2014} C. Zu, W.-B. Wang, L. He, W.-G. Zhang, C.-Y. Dai, F. Wang, L.-M. Duan,
	\enquote{Experimental realization of universal geometric quantum gates with solid-state spins,}
	Nature {\bf 514}, 72 (2014).
	%\bibitem{Zu_Nature2014} Zu, C. et al. Experimental realization of universal geometric
	%quantum gates with solid-state spins. Nature 514, 72-75 (2014). [27]
	
	\bibitem{Arroyo-Camejo_NatCommun2014} S. Arroyo-Camejo, A. Lazariev, S. W. Hell, and G. Balasubramanian,
	\enquote{Room temperature high-fidelity holonomic single-qubit gate on a solid-state spin,}
	Nat. Commun. {\bf 5}, 4870 (2014).
	%\bibitem{Arroyo-Camejo_NatCommun2014} Arroyo-Camejo, S., Lazariev, A., Hell, S. W. \& Balasubramanian,
	%G. Room temperature high-fidelity holonomic single-qubit gate on a
	%solid-state spin. Nat. Commun. 5, 4870 (2014). [28] 
	
	\bibitem{Abdumalikov_Nature2013} A. A. Abdumalikov Jr, J. M. Fink, K. Juliusson, M. Pechal, S. Berger,
	A. Wallraff, and S. Filipp,
	\enquote{Experimental realization of non-Abelian nonadiabatic geometric gates,}
	Nature (London) {\bf 496}, 482 (2013).
	%\bibitem{Abdumalikov_Nature2013} Abdumalikov, A. A. Jr. et al. Experimental realization of
	%non-Abelian nonadiabatic geometric gates. Nature 496, 482-485 (2013). [29] 
	
	\bibitem{Xu_PRL2018} Y. Xu, W. Cai, Y. Ma, X. Mu, L. Hu, Tao Chen, H. Wang, Y.-P. Song,
	Z.-Y. Xue, Z.-Q. Yin, L. Sun,
	\enquote{Single-loop realization of arbitrary non-adiabatic holonomic single-qubit quantum gates in a superconducting circuit,}
	Phys. Rev. Lett. {\bf 121}, 110501 (2018).
	%\bibitem{Xu_PRL2018} Y. Xu, W. Cai, Y. Ma, X. Mu, L. Hu, Tao Chen, H. Wang, Y.P. Song,
	%Zheng-Yuan Xue, Zhang-qi Yin, L. Sun, Single-loop realization of arbitrary
	%non-adiabatic holonomic single-qubit quantum gates in a superconducting
	%circuit, Phys. Rev. Lett. 121, 110501 (2018). [30] 
	
	\bibitem{Yan_PRL2019} T. Yan, B.-J. Liu, K. Xu, C. Song, S. Liu, Z.
	Zhang, H. Deng, Z. Yan, H. Rong, K. Huang, M.-H. Yung,
	Y. Chen, and D. Yu,
	\enquote{Experimental realization of nonadiabatic shortcut to non-Abelian geometric gates,}
	Phys. Rev. Lett. {\bf 122}, 080501 (2019).
	%\bibitem{Yang_PRL2019} Tongxing Yan, Bao-Jie Liu, Kai Xu, Chao Song, Song Liu, Zhensheng
	%Zhang, Hui Deng, Zhiguang Yan, Hao Rong, Keqiang Huang, Man-Hong Yung,
	%Yuanzhen Chen, and Dapeng Yu, Experimental Realization of Nonadiabatic
	%Shortcut to Non-Abelian Geometric Gates, Phys. Rev. Lett. 122, 080501 (2019). [31] 
	
	\bibitem{Zhang_arXiv1811} Z. Zhang, P.-Z. Zhao, T. Wang, L. Xiang, Z.
	Jia, P. Duan, D.-M. Tong, Y. Yin, and G. Guo,
	\enquote{Single-shot realization of nonadiabatic holonomic gates with a superconducting Xmon qutrit,}
	New J. Phys. {\bf 21}, 073024 (2019).
	%\bibitem{Zhang_arXiv1811} Zhenxing Zhang, P. Z. Zhao, Tenghui Wang, Liang Xiang, Zhilong
	%Jia, Peng Duan, D. M. Tong, Yi Yin, and Guoping Guo, Single-shot realization
	%of nonadiabatic holonomic gates with a superconducting Xmon qutrit,
	%arXiv:1811.06252. [32] 
	
	\bibitem{You_PhysicsToday2005} J. Q. You and F. Nori,
	\enquote{Superconducting circuits and quantum
		information,} Phys. Today {\bf 58}, 42 (2005). %[33] 
	
	\bibitem{Egger_PRApplied2019} D. J. Egger, M. Ganzhorn, G. Salis, A. Fuhrer, P. Mueller, P. K.
	Barkoutsos, N. Moll, I. Tavernelli, and S. Filipp,
	\enquote{Entanglement generation in superconducting qubits using holonomic operations,}
	Phys. Rev. Applied {\bf 11}, 014017 (2019).
	%\bibitem{Egger_PRApplied2019} D. J. Egger, M. Ganzhorn, G. Salis, A. Fuhrer, P. Mueller, P. K.
	%Barkoutsos, N. Moll, I. Tavernelli, and S. Filipp, Phys. Rev. Applied 11,
	%014017 (2019). [34]
	
	\bibitem{Han_arXiv2004} Z. Han, Y. Dong, B. Liu, X. Yang, S. Song, L. Qiu, D. Li, J. Chu, W. Zheng, J. Xu, T. Huang, Z. Wang, X. Yu, X. Tan, D. Lan, M.-H. Yung, and Y. Yu,
	\enquote{Experimental realization of universal time-optimal non-Abelian geometric gates,}
	arXiv:2004.10364 (2020).
	%\bibitem{Han_arXiv2004} Z. Han, Y. Dong, B. Liu, X. Yang, S. Song, L. Qiu, D. Li, J. Chu,
	%W. Zheng, J. Xu, T. Huang, Z.Wang, X. Yu, X. Tan, D. Lan, M.-H. Yung, and Y.
	%Yu, \textquotedblleft Experimental realization of universal time-optimal
	%non-Abelian geometric gates,\textquotedblright\ arXiv:2004.10364 (2020). [35] 
	
	\bibitem{Buluta_RopiP2011} I. Buluta, S. Ashhab, and F. Nori,
	\enquote{Natural and artificial atoms for quantum computation,} Reports on Progress in Physics {\bf 74}, 104401 (2011). %[36] 
	
	\bibitem{Ning_PRL2019} W. Ning, X.-J. Huang, P.-R. Han, H. Li, H. Deng, Z.-B. Yang, Z.-R. Zhong,
	Y. Xia, K. Xu, D. Zheng, and S.-B. Zheng,
	\enquote{Deterministic entanglement swapping in a superconducting circuit,}
	Phys. Rev. Lett. {\bf123}, 060502 (2019).
	%\bibitem{Ning_PRL2019} W. Ning, X.-J. Huang, P.-R. Han, H. Li, H. Deng, Z.-B. Yang,
	%Z.-R. Zhong, Y. Xia, K. Xu, D. Zheng, and S.-B. Zheng, \textquotedblleft
	%Deterministic entanglement swapping in a superconducting
	%circuit,\textquotedblright\ Phys. Rev. Lett. 123, 060502 (2019). [37]
	
	\bibitem{Yang_npj_Quantum Information2021} Z.-B. Yang, P.-R. Han, X.-J. Huang, W. Ning, H. Li, K. Xu, D. Zheng, H. Fan, and S.-B. Zheng,
	\enquote{Experimental demonstration of entanglement-enabled universal quantum cloning in a circuit,}
	npj Quantum Inf {\bf7}, 44 (2021).
	
	
	\bibitem{Yurtalan_arXiv2020} M. A. Yurtalan, J. Shi, G. J. K. Flatt, and A. Lupascu,
	\enquote{Characterization of multi-level dynamics and decoherence in a high-anharmonicity capacitively shunted flux circuit,}
	arxiv:2008.00593 (2020).
	%\bibitem{Yurtalan_arXiv2020} M. A. Yurtalan, J. Shi, G. J. K. Flatt, and A. Lupascu,
	%\textquotedblleft Characterization of multi-level dynamics and decoherence
	%in a high-anharmonicity capacitively shunted flux
	%circuit,\textquotedblright\ arxiv:2008.00593 (2020). [38] 
	
	\bibitem{You_PRB2007} J. Q. You, X. Hu, S. Ashhab and F. Nori,
	\enquote{Low-decoherence flux qubit,} Phys. Rev. B {\bf 75}, 140515(R) (2007). %[39]
	
	\bibitem{Fowler_PRA2012} A. G. Fowler, M. Mariantoni, J. M. Martinis, and A. N. Cleland,
	\enquote{Surface codes: Towards practical large-scale quantum computation,}
	Phys. Rev. A {\bf 86}, 032324 (2012)
	%\bibitem{Fowler_PRA2012} A. G. Fowler, M. Mariantoni, J. M. Martinis, and A. N. Cleland,
	%\textquotedblleft Surface codes: Towards practical large-scale quantum
	%computation,\textquotedblright\ Phys. Rev. A 86, 032324 (2012). [40]
	
	\bibitem{Barends_Nature2014} R. Barends, J. Kelly, A. Megrant, A. Veitia, D. Sank, E. Jeffrey,
	T. C. White, J. Mutus, A. G. Fowler, B. Campbell, Y. Chen, Z. Chen, B. Chiaro, A. Dunsworth,
	C. Neill, P. O'Malley, P. Roushan, A. Vainsencher, J. Wenner, A. N. Korotkov, A. N. Cleland, and
	J. M. Martinis,
	\enquote{Superconducting quantum circuits at the surface code threshold for fault tolerance,}
	Nature (London) {\bf 508}, 500 (2014).
	%\bibitem{Barends_Nature2014} R. Barends, J. Kelly, A. Megrant, A. Veitia, D. Sank, E. Jeffrey,
	%T. C. White, J. Mutus, A. G. Fowler, B. Campbell, Y. Chen, Z. Chen, B.
	%Chiaro, A. Dunsworth, C. Neill, P. O'Malley, P. Roushan, A. Vainsencher, J.
	%Wenner, A. N. Korotkov, A. N. Cleland, and J. M. Martinis, \textquotedblleft
	%Superconducting quantum circuits at the surface code threshold for
	%fault-tolerance,\textquotedblright\ Nature (London) 508, 500 (2014). [41]
	
	\bibitem{Barends_Nature2015} R. Barends, A. G. Fowler, A. Megrant, E. Jeffrey, T. C. White, D. Sank, J. Y. Mutus, B. Campbell, Yu Chen, Z. Chen, B. Chiaro, A. Dunsworth, I.-C. Hoi, C. Neill, P. J. J. O'Malley, C. Quintana, P. Roushan, A. Vainsencher, J. Wenner, A. N. Cleland, and John M. Martinis,
	\enquote{State preservation by repetitive error detection in a superconducting quantum circuit,}
	Nature (London) {\bf 519}, 66 (2015).
	%\bibitem{Barends_Nature2015} R. Barends, A. G. Fowler, A. Megrant, E. Jeffrey, T. C. White, D.
	%Sank, J. Y. Mutus, B. Campbell, Yu Chen, Z. Chen, B. Chiaro, A. Dunsworth,
	%I.-C. Hoi, C. Neill, P. J. J. O'Malley, C. Quintana, P. Roushan, A.
	%Vainsencher, J. Wenner, A. N. Cleland, and John M. Martinis,
	%\textquotedblleft State preservation by repetitive error detection in a
	%superconducting quantum circuit,\textquotedblright\ Nature (London) 519, 66
	%(2015).  [42] 
	
\end{thebibliography}
\end{document}

\maketitle

\section{Device parameters}
The device used in our experiment is similar to the ones reported in Refs. \cite{song_nc2017,ning_prl2019,Yang_npj_Quantum Information2021}, it possesses 
five superconducting Xmon qutrits, whose basis states are denoted as $\left|g\right\rangle$, $\left|e\right\rangle$ and $\left|f\right\rangle$, as shown in Fig. 1 of the main text. These qutrits are capacitively connected to a bus resonator, whose rare frequency is measured as $5.584$ GHz 
when all the qutrits are staying in the ground state $\vert g\rangle$ at their respective idle frequencies $\omega_j/2\pi$.  
Each qutrit's frequency is flexibly adjusted and thus can be controlled to couple to the bus resonator in a dispersive or resonant manner,   
that induces virtual-photon-mediated \cite{ning_prl2019,10qGHZ} or real-photon-mediated qubit-qubit coupling \cite{erik_np2012}. The coupling strength $\lambda_j$ between each qutrit and the 
resonator through the $\vert g\rangle\longleftrightarrow \vert e\rangle$ transition is measured through qutrit-resonator vacuum Rabi swap, while 
keeping the qutrit's $\vert e\rangle\longleftrightarrow \vert f\rangle$ transition decoupled as the qutrit's anharmonicity $\alpha_j$ is much larger than $\lambda_j$. 
The device is kept inside a dilution refrigerator with a base temperature below 20 mK. 
We pick up two qutrits, which are labelled as Q$_1$ and Q$_2$, for our implementation. 
The related parameters including the qutrit states' coherence times and readout fidelities are characterized and listed in Table~\ref{table1}.    
As the computation information is encoded in $\left|g\right\rangle$ and $\left|f\right\rangle$ ($\left|e\right\rangle$ as the auxiliary state), we thus refer to the qutrits as qubits.       

\begin{table*}[htbp]
	\centering
	\caption{\label{table1} \textbf{Qubits characteristics.}}
	\resizebox{\textwidth}{10mm}{ 
	\begin{tabular}{cccccccccccc}
		%\centering
		\hline
		\hline
		&$\omega_{j}/2\pi$ (GHz)&$T_{1,j}^{\vert e\rangle}$ ($\mu$s)&$T_{2,j}^{*,{\vert e\rangle}}$ ($\mu$s)&$T_{2,j}^{\textrm{SE},{\vert e\rangle}}$ ($\mu$s)& $T_{1,j}^{\vert f\rangle}$ ($\mu$s)&$T_{2,j}^{*,{\vert f\rangle}}$ ($\mu$s)&$\alpha_j/2\pi$&$\lambda_j/2\pi$ (MHz) &$F_j^{\vert g\rangle}$&$F_j^{\vert e\rangle}$&$F_j^{\vert f\rangle}$\\
		\hline
		$Q_1$&5.47&23.9&2.7&7.6&13.0&2.1&242&20.8&0.96&0.84&0.87\\
		$Q_2$&5.34&15.9&2.1&8.5&10.7&1.5&249&19.9&0.98&0.87&0.89\\
		%$Q_3$&5.366&24.0&2.0&6.4&55.5&20.0&315&0.961&0.919\\
		%$Q_4$&5.421&18.1&2.0&7.6&45.2&19.4&313&0.979&0.822\\
		\hline
		\hline
	\end{tabular}}
		\justifying
			The idle frequency of Q$_j$ is $\omega_{j}/2\pi$, where single-qubit rotation pulses and tomographic pulses are applied. Here $T_{1,j}^{\vert k\rangle}$, $T_{2,j}^{*,\vert k\rangle}$ and $T_{2,j}^{\textrm{SE},\vert k\rangle}$ ($k=e,f$) are respectively the energy relaxation time, the Ramsey dephasing time and the spin-echo dephasing time of Q$_j$'s state $\vert k\rangle$ measured at the idle point. In addition, $\alpha_j $ is the qubit's anharmonicity and 
        	$\lambda_j$ is the coupling strength between Q$_j$ and the bus resonator. The probability of detecting Q$_j$ in state $\vert k\rangle$ when it is prepared in state $\vert k\rangle$ is $F_{j}^{\vert k\rangle}$. The $I$-$Q$ data to differentiate these basis states are plotted in Fig. \ref{fS7}.
		\justifying	 
\end{table*}

\begin{figure}[htbp] 
	\centering
	\includegraphics[width=5.0in]{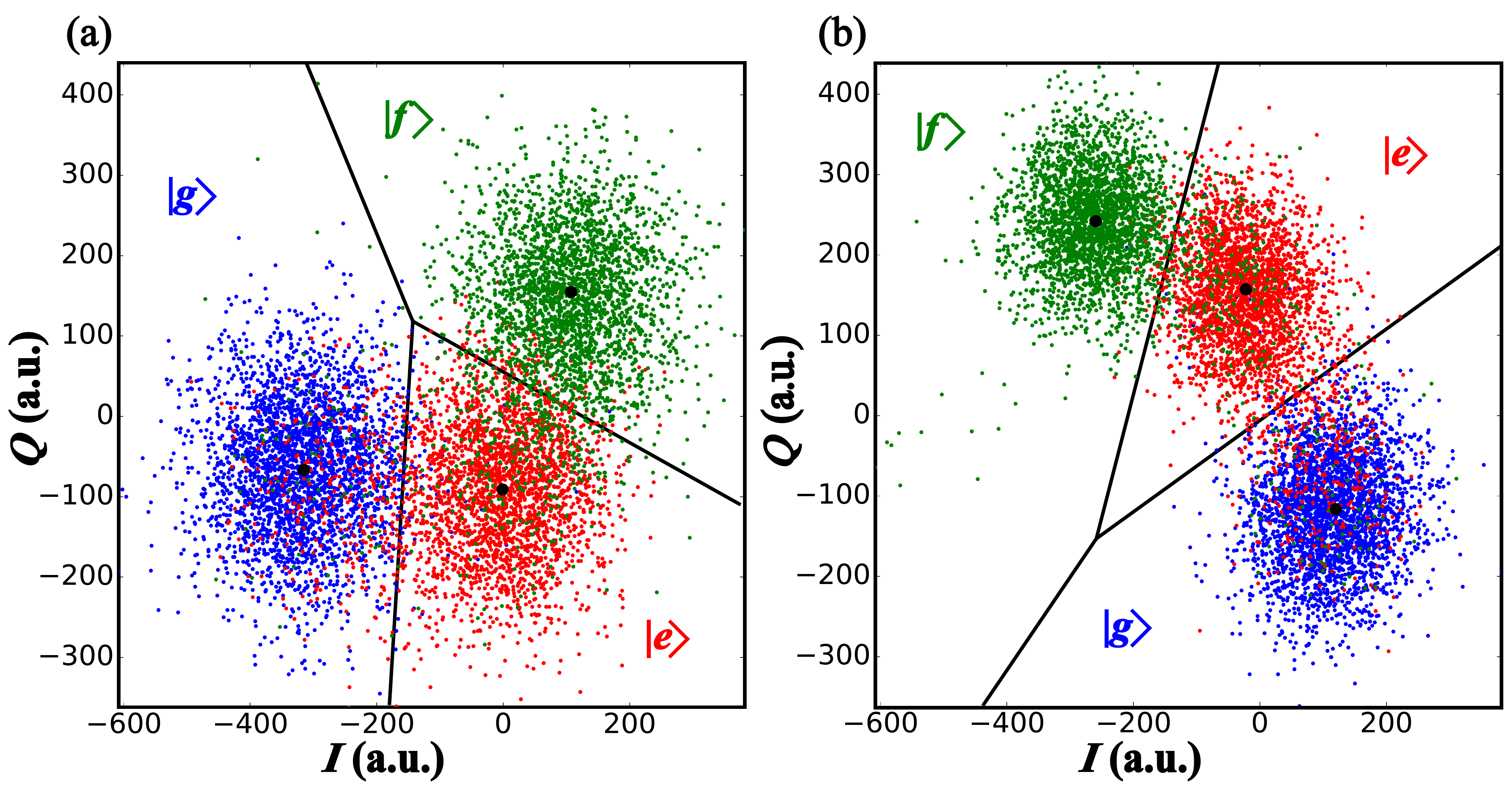}
	\caption{ The measured $I$-$Q$ values when each qubit is prepared in $\vert g\rangle$ (blue), $\vert e\rangle$ (red) and $\vert f\rangle$ (green) state. (a) The $I$-$Q$ data of the control qubit Q$_1$. (b) The $I$-$Q$ data of the target qubit Q$_2$.}
	\label{fS7}
\end{figure}

\section{Stark shifts induced by the off-resonant coupling}
%The realization of the non-Abelian geometric controlled-Not gate is based on the holonomic manipulation of the system through a pair of classical fields, which induce the system state to evolve across the specific dressed state, whose energy level is %determined by the system's excitation number and the number of qubits being initialized in $\vert g\rangle$.
\begin{figure}[htbp] 
	\centering
	\includegraphics[width=3.3in]{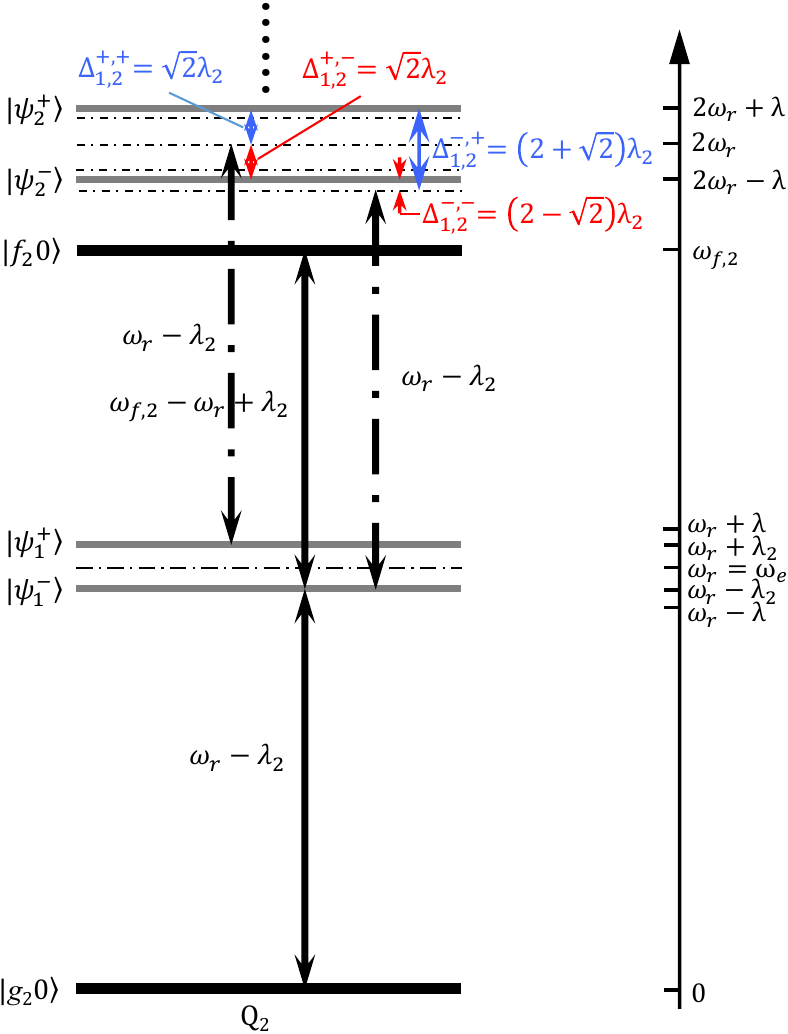}
	\caption{Schematic diagram of the off-resonant couplings between the driving fields and the transitions $\vert \psi_1^{\pm}\rangle \longleftrightarrow \vert \psi_2^{\pm}\rangle$. These lead to the energy shift $-2\hbar\Omega_{ge}^2/\lambda_2$, for $\vert \psi_1^-\rangle$.}
	\label{fS1}
\end{figure}
 
\begin{figure}[htbp] 
	\centering
	\includegraphics[width=3.5in]{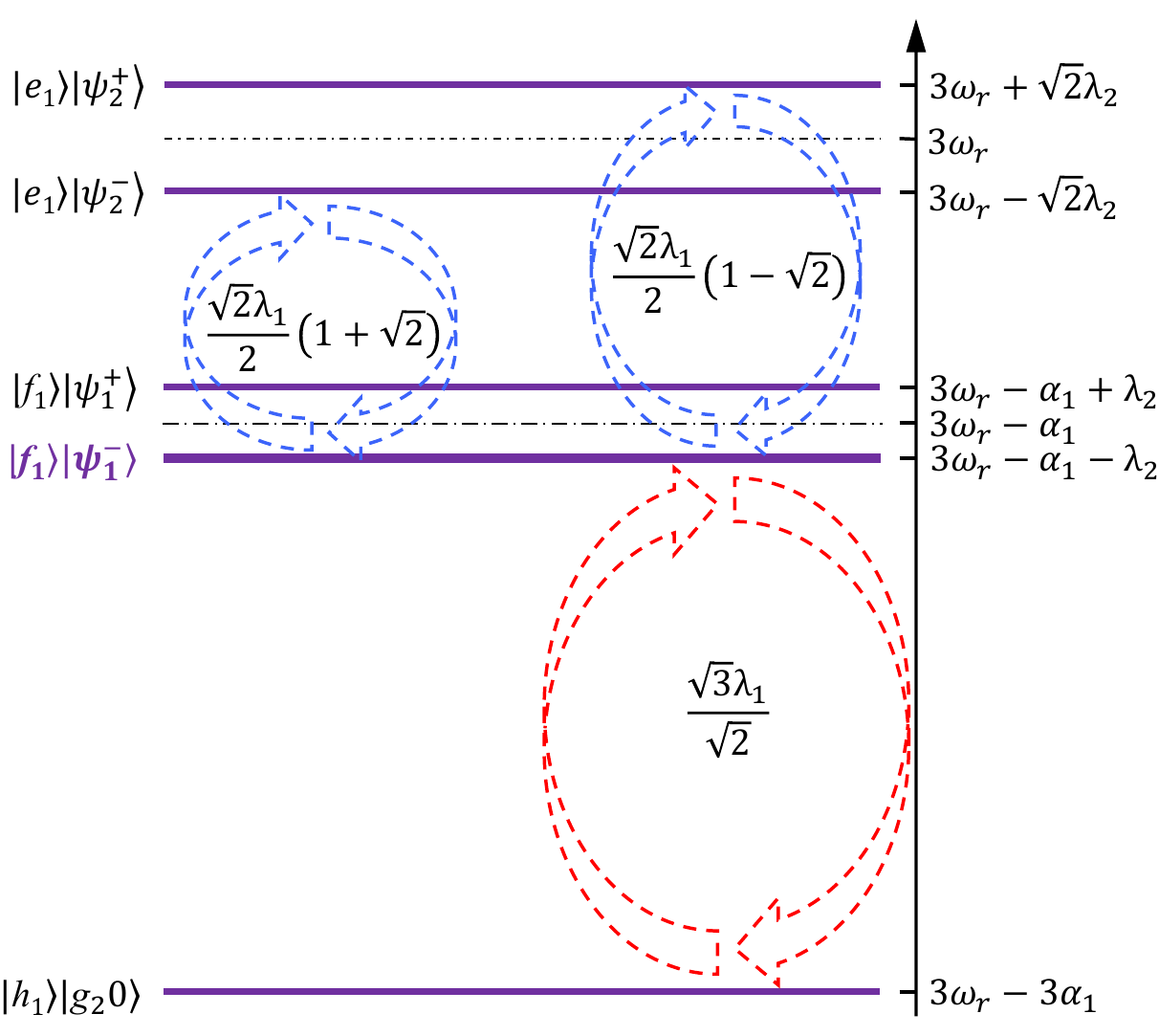}
	\caption{Schematic diagram of the off-resonant couplings between the resonator photons and the transitions $\vert f_1\rangle\vert \psi_1^{\pm}\rangle \longleftrightarrow \vert h_1\rangle\vert \psi_2^{\pm}\rangle$ and $\vert f_1\rangle\vert \psi_1^{\pm}\rangle \longleftrightarrow \vert e_1\rangle\vert \psi_2^{\pm}\rangle$. The photon-induced Stark shifts for $\vert \psi_1^{\pm}\rangle$ are $\hbar\delta_2\simeq-9\lambda_1^2/4\alpha_1$, approximately.}
	\label{fS2}
\end{figure}

When the system is initially in $\vert f_1g_20\rangle$, the control qubit does not interact with the resonator, while the target qubit strongly couples with the resonator. Such a strong coupling produces the dressed states $\vert \psi_n^{\pm}\rangle$, with the corresponding eigenenergies $\hbar(n\omega_r\pm\sqrt{n}\lambda_2)$. The two microwave fields with the angular frequencies 
\begin{equation}
    \omega_{d1}=\omega_r-\lambda_2
\end{equation}
and
\begin{equation}
    \omega_{d2}=\omega_f-\omega_r+\lambda_2
\end{equation}
resonantly drive the two transitions $\vert g_20\rangle \longleftrightarrow \vert \psi_1^{-}\rangle$ and $\vert f_20\rangle \longleftrightarrow \vert \psi_1^{-}\rangle$, respectively, as depicted in Fig. \ref{fS1}. As the Rabi frequencies $\Omega_{ge}$ and $\Omega_{ef}$ of the two driving fields are much smaller than the qubit-resonator coupling strength $\lambda_2$, the two fields cannot drive the transition from $\vert \psi_1^{-}\rangle$ to $\vert \psi_2^{\pm}\rangle$, whose energy gaps are $\hbar[\omega_r +(1\pm\sqrt{2})\lambda_2]$, largely detuned from the two fields by the amount of 
\begin{equation}
    \Delta_{1,2,d1}^{-,\pm}=(2\pm\sqrt{2})\lambda_2
\end{equation} 
and 
\begin{equation}
    \Delta_{1,2,d2}^{-,\pm}=2\omega_r-\omega_f\pm\sqrt{2}\lambda_2,
\end{equation} 
respectively. However, these off-resonant couplings shift the energy level of $\vert \psi_1^-\rangle$ by about $-\hbar \delta_1$, with $\delta_1=\Omega_{ge}^2/(2-\sqrt{2})\lambda_2+\Omega_{ge}^2/(2+\sqrt{2})\lambda_2\equiv 2\Omega_{ge}^2/\lambda_2$. 
Besides, off-resonant couplings from $\vert f_1\rangle \vert \psi_1^-\rangle$ to $\vert h_1\rangle \vert g_20\rangle$ and $\vert e_1\rangle \vert \psi_2^{\pm}\rangle$ through the resonator photon also lead to energy shifts (see Fig. \ref{fS2}), which are 
\begin{equation}
	\hbar\delta_{2,1}=\hbar(\frac{\sqrt{3}\lambda_1}{\sqrt{2}})^2/(2\alpha_1-\lambda_2),
\end{equation} 
and
\begin{equation}
     \hbar\delta_{2,2}^{\pm}=-\hbar[\frac{\sqrt{2}\lambda_1}{2}(1\mp \sqrt{2})]^2/[\alpha_1+(1\pm \sqrt{2})\lambda_2],
\end{equation} 
respectively, three summing up to about $\hbar\delta_2\simeq-9\lambda_1^2/4\alpha_1$.  

Note that the two fields also cannot drive the transition from $\vert \psi_1^+\rangle \longleftrightarrow \vert \psi_2^{\pm}\rangle$, for which the energy gaps are $\hbar[\omega_r-(1\mp\sqrt{2})\lambda_2]$, largely detuned from the two fields by
\begin{equation}
	\Delta_{1,2,d1}^{+,\pm}=\pm\sqrt{2}\lambda_2
\end{equation}
and  
\begin{equation}\Delta_{1,2,d2}^{+,\pm}=2\omega_r-\omega_f-(2\mp\sqrt{2})\lambda_2,
\end{equation} 
respectively.
Though the different off-resonant couplings lead to respective energy shifts to $\vert \psi_1^+\rangle$, such energy shifts are symmetric and thus neutralize to keep $\vert \psi_1^+\rangle$ almost constant. Note also that off-resonant couplings from $\vert f_1\rangle \vert \psi_1^+\rangle$ to $\vert h_1\rangle \vert g_20\rangle$ and $\vert e_1\rangle \vert \psi_2^{\pm}\rangle$ through the resonator photon 	lead to energy shifts, which are \begin{equation}\hbar\delta_{2,1}^\prime=\hbar(\frac{\sqrt{3}\lambda_1}{\sqrt{2}})^2/(2\alpha_1+\lambda_2)
\end{equation} 
and
\begin{equation}\hbar\delta_{2,2}^{\pm \prime}=-\hbar[\frac{\sqrt{2}\lambda_1}{2}(1\pm \sqrt{2})]^2/[\alpha_1-(1\mp \sqrt{2})\lambda_2],
\end{equation}
respectively, adding up to also about $\hbar\delta_2\simeq-9\lambda_1^2/4\alpha_1$.

\begin{figure}[htbp] 
	\centering
	\includegraphics[width=5.0in]{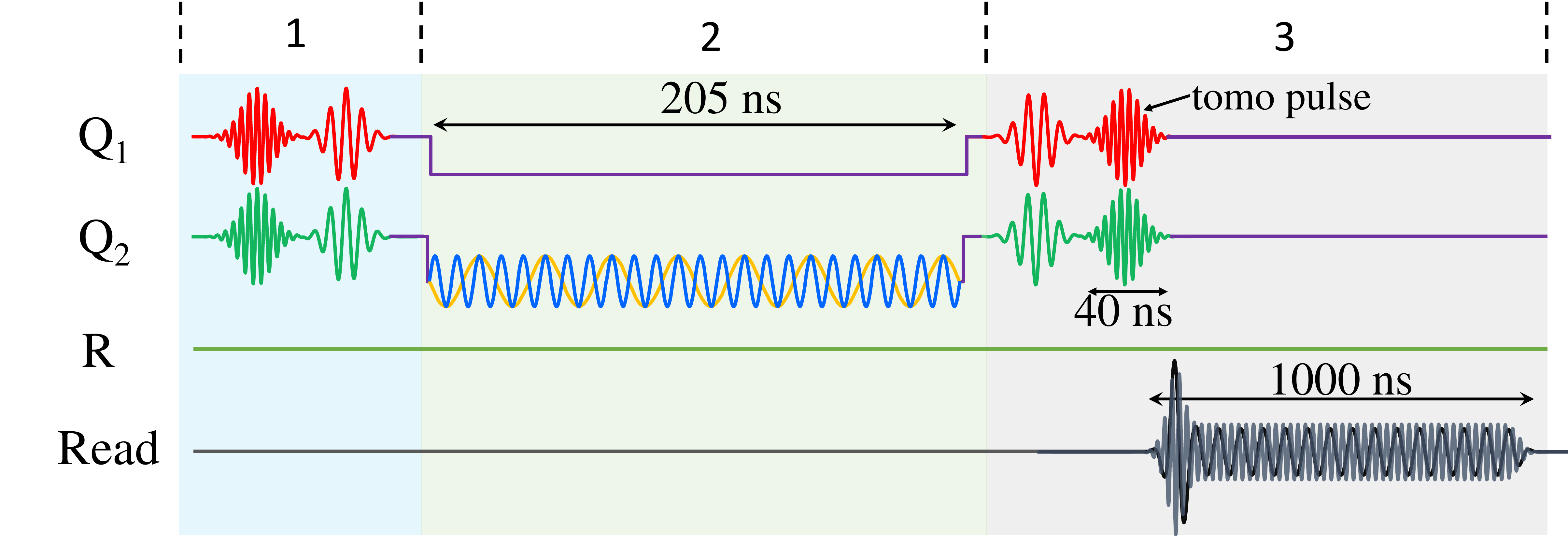}
	\caption{ The experimental sequence, which consists of three steps: initial state preparation, gate operation and quantum state tomography. The initial state is created by applying microwave pulses with a Gaussian envelop at the idle points. Then, in the second step, qubit frequencies are tuned by rectangular waves to be near (Q$_1$) or on resonance (Q$_2$) with the resonator. Q$_2$ is subjected to a two-tone microwave pulse with a flattop envelop during the interaction time which lasts about 205 ns. In the third step, tomographic operations are executed before the two-qubit joint readout.}
	\label{fS3}
\end{figure}

\begin{figure}[htbp] 
	\centering
	\includegraphics[width=5.0in]{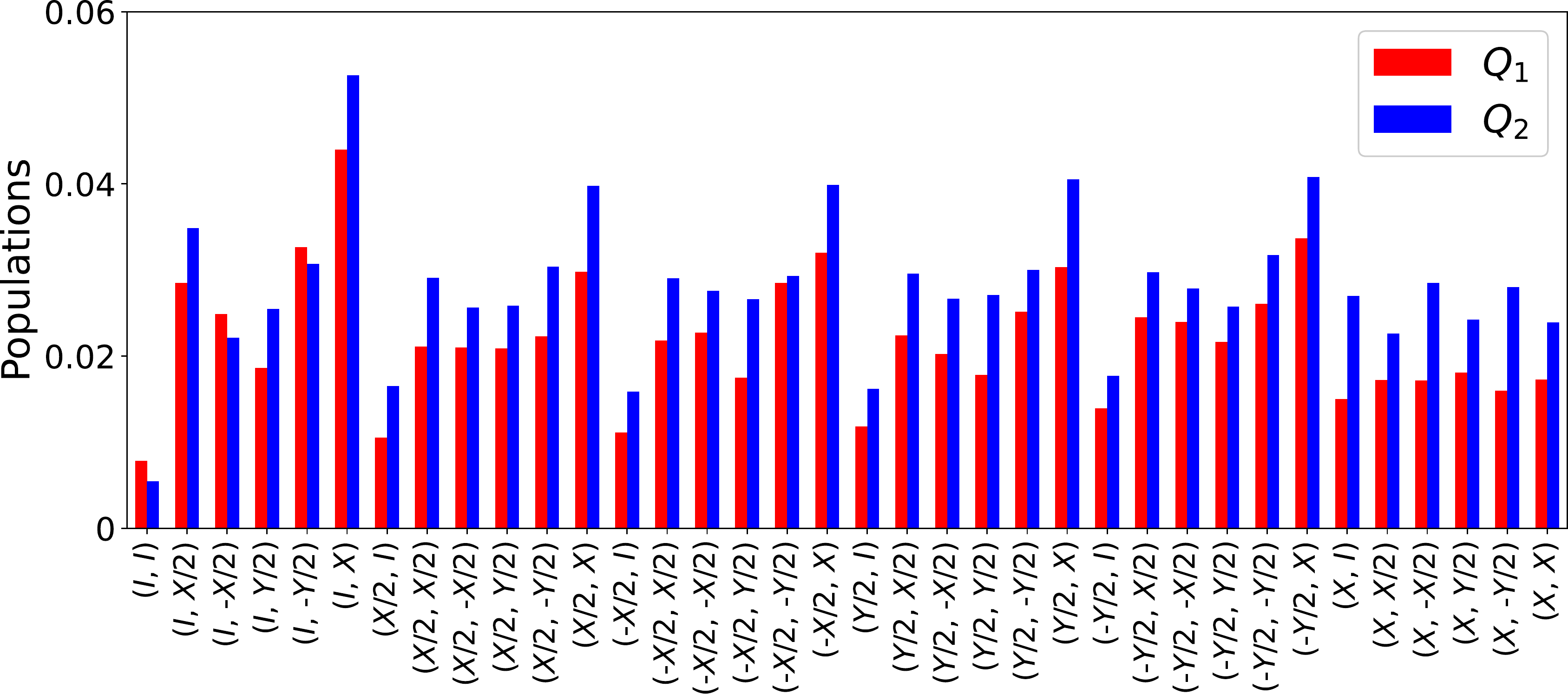}
	\caption{The probabilities of leaking to the $\vert e\rangle$ state for each qubit. The probabilities are measured after the gate sequence is finished for all 36 input states. The labels in the $x$-axis represent the single qubit rotations used to prepare the initial states.}
	\label{fS6}
\end{figure}

\section{Experimental sequence of the holonomic gate}
Figure \ref{fS3} shows the experimental sequence, which is divide into three steps. Firstly, two successive microwave pulses are imposed on each qubit at their idle points to prepare the initial state. The first pulse with the frequency of $\omega_j/2\pi$ realizes the $\vert g\rangle \leftrightarrow \vert e\rangle$ rotation while the second pulse with the frequency of $(\omega_j-\alpha_j)/2\pi$  is a flip operation between $\vert e\rangle$ and $\vert f\rangle$ state, known as a $e$-$f$ $\pi$ rotation. After the initial state preparation, rectangular pulses are applied to open the qubit-resonator interaction for a time of about 205 ns. The control qubit Q$_1$ is biased to an optimized point close to $\omega_r$, while the target qubit Q$_2$ stays on resonance with the resonator when a two-tone microwave pulse with the angular frequencies $\omega_r-\lambda_2-\delta_1-\delta_2$ and $\omega_{f,2}-\omega_r+\lambda_2+\delta_1+\delta_2$ are applied on Q$_2$. Finally, the qubits are brought back to their idle points for quantum state tomography. To extract the density matrix,  we use three tomographic operations \{$I$, $X/2$, $Y/2$\} which are executed at the $\{\vert g\rangle, \vert e\rangle\}$ space after an $e$-$f$ $\pi$ rotation for each qubit, as can be seen in the third step of the sequence. For each tomographic operation, we perform the two-qubit joint readout by applying a two-tone measurement pulse to the transmission line, yielding the probabilities of the two-qubit basic states $\{\vert g\rangle,\vert e\rangle,\vert f\rangle\}^{\otimes 2}$. As only probabilities of $\vert g\rangle$ and $\vert f\rangle$ state are used for post analysis, the extracted density matrices have a trace value of smaller than 1, which indicates a leakage to the $\vert e\rangle$ state. In Fig. \ref{fS6} we plot the measured leakage probability in $\vert e\rangle$ state for each qubit.

\begin{figure}[htbp] 
	\centering
	\includegraphics[width=5.0in]{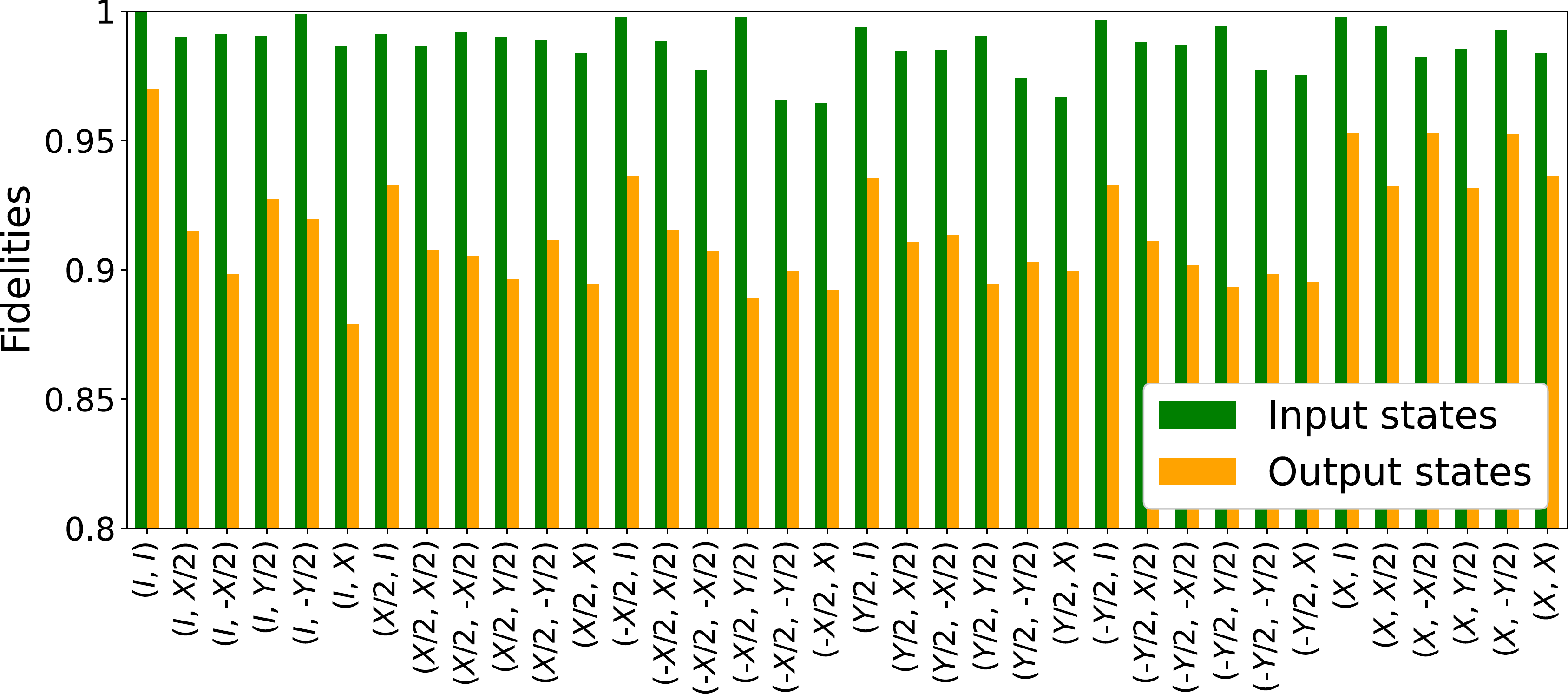}
	\caption{ The experimental fidelities of the input and output states, A total of 36 input and output states are used to perform the quantum process tomography. The labels in the $x$-axis represent the single qubit rotations used to prepare the initial states.}
	\label{fS5}
\end{figure}

\section{Quantum process tomography}
Quantum process tomography is executed by performing the state tomography for totally 36 input and corresponding output states after the gate sequence are applied. The input states are two-qubit product states 
\begin{equation}
    \{\vert g\rangle, \frac{1}{\sqrt{2}}(\vert g\rangle-i\vert f\rangle),\frac{1}{\sqrt{2}} (\vert g\rangle+i\vert f\rangle), \frac{1}{\sqrt{2}}(\vert g\rangle+\vert f\rangle), \frac{1}{\sqrt{2}}(\vert g\rangle-\vert f\rangle), \vert f\rangle\}^{\otimes 2}
\end{equation}
which are produced by applying rotation pulses to each qubit. The mean fidelity characterized by quantum state tomography for all input states and output states are about $0.983\pm0.003$ and $0.915\pm0.008$, respectively, as shown in Fig. \ref{fS5}. The $\chi$-matrix can be extracted from these input and output states by utilizing the least square optimization method with the Hermitian and positive semidefinite constraints \cite{Neeley2010}. Note that we did not apply the constraint of unit trace for both the calculation of density matrix and $\chi$-matrix considering the leakage to non-computational states.

\iffalse
\begin{table*}[!htb]
	\centering
	\begin{tabular}{c|ccccc|ccccc}
		\hline
		\hline
		Nonlinearity, $\alpha/2\pi$ (GHz)&0.247&0.5&0.8&1.0&2.0&0.247&0.247&0.247&0.247&0.247\\
		Driving amplitude, $\Omega/2\pi$ (MHz)&2.3&2.3&2.3&2.3&2.3&2.3&1.8&1.5&1.0&0.5\\
		Gate time (ns)&209.5&220.4&220.1&223.5&223.8&209.5&281.0&360.5&532.5&1116.0\\
		$\chi$-fidelity with decoherence considered&0.9075&0.9321&0.9423&0.9437&0.9448&0.9075&0.9058&0.8915&0.8837&0.7985\\
		$\chi$-fidelity without decoherence considered&0.9415&0.9694&0.9797&0.9818&0.9839&0.9415&0.9506&0.9493&0.9668&0.9651\\
		
		\hline
		\hline
	\end{tabular}
\fi

\begin{table*}[!htb]
	\centering
	\caption{\label{table2} \textbf{Numerical results.} }
	\resizebox{\textwidth}{14mm}{ 
	\begin{tabular}{c|ccccc|ccccc|c}
		\hline
		\hline
		Nonlinearity, $\alpha_j/2\pi$ (GHz)&0.247&0.5&0.8&1.0&2.0&0.247&0.247&0.247&0.247&0.247&1.0\\
		Driving amplitude, $\Omega_{ge,ef}/2\pi$ (MHz)&2.3&2.3&2.3&2.3&2.3&2.3&1.8&1.5&1.0&0.5&1.5\\
		Gate time (ns)&209.5&220.4&220.1&223.5&223.8&209.5&281.0&360.5&532.5&1116.0&335.9\\
		$\chi$-fidelity with decoherence considered&0.908&0.932&0.942&0.944&0.945&0.908&0.906&0.892&0.884&0.799&0.935\\
		$\chi$-fidelity without decoherence considered&0.942&0.969&0.980&0.982&0.984&0.942&0.951&0.949&0.967&0.965&0.993\\
		
		\hline
		\hline
	\end{tabular}}
\justifying
{  The gate fidelities for different nonlinearities and driving amplitudes are obtained by optimizing the evolution time and qubit frequencies. The cases with decoherence adopt the $T_1$ values listed in Table~\ref{table1} and the pure dephasing times of about 40 $\mu s$. The dephasing time used here is estimated from the exponential fit of the Ramsey measurement data before 200 ns. The first column presents numerical data considering parameters of our experimental device, which shows a good agreement with the experimental results. The limitation of nonlinearity, restriction of driving amplitude and decoherence contribute gate errors of about 4.2\%, 2.3\% and 3.4\% respectively.}
\justifying
\end{table*}

\section{Gate error analysis}
We have performed numerical simulation to quantify the errors of our gate. The infidelity of our gate mainly comes from the imperfect decoupling between the microwave drive and the qubit and also the decoupling between the qubit and the resonator. For example, when the control qubit Q$_1$ is in $\vert g\rangle$ state, the detuning between the microwave drive and the dressed state energy level is not large enough to decouple them, which induces a small transition from $\vert g_1,g_2,0\rangle$ and $\vert g_1,f_2,0\rangle$ to $\vert \Phi_1^0\rangle$ and $\vert \Phi_1^{\pm}\rangle$, leading to a leakage error. Lowering the drive amplitude can effectively reduce this leakage error, but will extend the evolution time and as a result increase the decoherence error. In addition, the nonlinearities $\alpha_j$ need to be larger to better decouple the control qubit Q$_1$ from the resonator and the target qubit when Q$_1$ is prepared in $\vert f\rangle$ state during the gate operation.

Considering all these factors, in Table~\ref{table2}, we numerically calculate the $\chi$-fidelities for different nonlinearities and driving amplitudes. When decoherence is neglected, increasing the nonlinearity or decreasing the driving amplitude can both improve the fidelity, as shown in the last column of Table~\ref{table2}. The CNOT gate with $\chi$-fidelity larger than 0.99 can be realized by use of qubits with good coherence, provided the nonlinearity reaches $2\pi\times1.0$ GHz and the driving amplitude reduces to $2\pi\times1.5$ MHz. However, small driving amplitude requires long evolution time, which leads to more decoherence error. For a gate time of about 200 ns, the decoherence contributes about 3.4\% of the total error, as shown in the table. To achieve a short evolution time, both the nonlinearity and coupling strength need to be enlarged. Our further numerical simulations show that, given parameters ($\alpha_j/2\pi=-3.69$ GHz, $\lambda_j/2\pi=110$ MHz, $\Omega_{ge,ef}/2\pi=5.9$ MHz, $T_{1,j}$=60 $\mu s$, $T_{2,j}^*$=86 $\mu s$) which are accessible in recent superconducting qubits~\cite{Yuan2020,petar2020,Yurtalan2020}, the CNOT gate with an optimized operation time of 87 ns yields an $\chi$-fidelity of 0.991, indicating the potential of our scheme in high-fidelity quantum operations.

\begin{figure}[htbp] 
	\centering
	\includegraphics[width=3.5in]{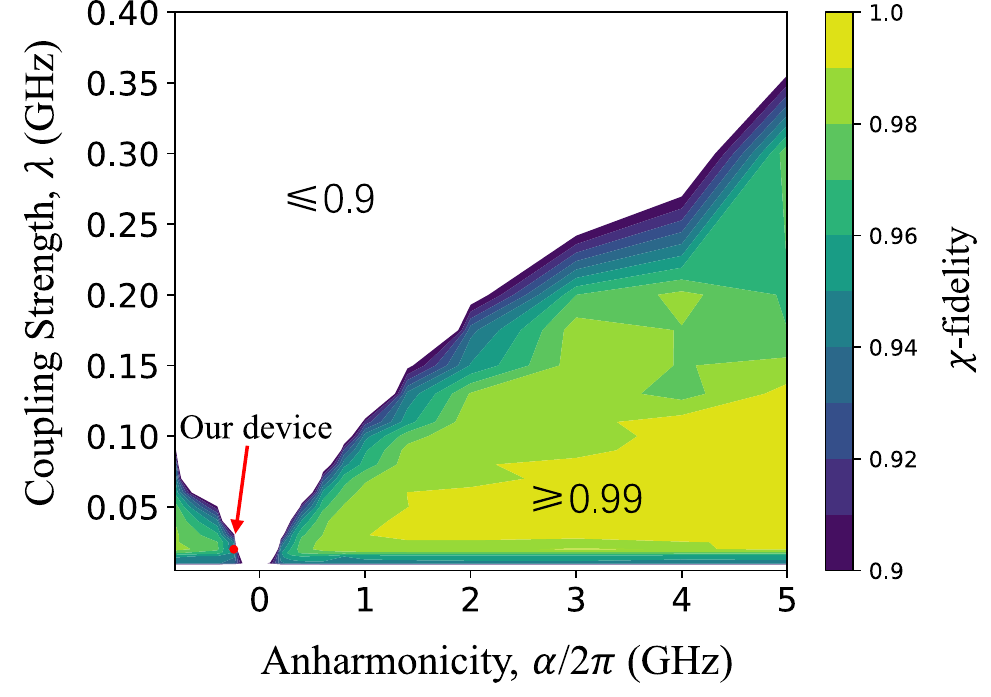}
	\caption{Numerical process fidelities of the CNOT gate by varying the anharmonicity $\alpha_j$ and coupling strength $\lambda_j$. Here w{\tiny }e have set $\alpha\equiv -\alpha_j$, $\lambda\equiv\lambda_j$ ($j=1,2$). The simulation results are obtained by optimizing the evolution time and qubit frequencies without considering decoherence. The red dot shows the position of our current device.}
	\label{f2D}
\end{figure}

\section{Additional numerical simulation}
We have further performed numerical simulation by sweeping both the anharmonicity $\alpha_j$ and qubit-resonator coupling strength $\lambda_j$. For simplicity, the drive amplitude is fixed to be $\Omega_{ge,ef}/2\pi=$2.0 MHz, similar to that used in our experiment. Fig.~\ref{f2D} plots the numerical process fidelities $\chi$ of the CNOT gate in parameter space, where the yellow region meets the threshold for surface code.
   
%\bibliographystyle{plain}
%\bibliography{enbib}